\documentclass[smallextended]{svjour3}       

\usepackage[authoryear]{natbib}
\setcitestyle{authoryear,open={(},close={)},citesep={;},aysep={,},yysep={,},maxnames=4}
\usepackage{url}

\smartqed  
\usepackage{graphicx}
\usepackage{comment}
\usepackage{amssymb}
\usepackage{ifthen}
\usepackage{paralist, tabularx}
\usepackage{listings}
\usepackage{tabularx}

\usepackage{booktabs}
\usepackage{siunitx}
\usepackage{caption}
\usepackage{multirow}
\usepackage{tabularx}
\usepackage{comment}

\usepackage{tcolorbox}

\usepackage{amssymb}
\usepackage[utf8]{inputenc}
\usepackage{amsmath}
\usepackage{multirow, listings, xcolor}
\usepackage{tcolorbox}
\usepackage{hyperref}
\usepackage{longtable}
\usepackage{booktabs}
\usepackage{rotating}
\usepackage{float}
\usepackage{pdflscape}
\usepackage[T1]{fontenc}
\newcommand{\gptmini}{GPT-4o-mini}

\tcbset{
  myinstructionbox/.style={
    colback=blue!5!white,
    colframe=blue!75!black,
    fonttitle=\bfseries,
    title=Instructions,
    sharp corners,
    boxrule=0.8pt,
    arc=4pt,
    left=6pt,
    right=6pt,
    top=6pt,
    bottom=6pt,
  }
}

\usepackage{listings}
\usepackage{xcolor}
\lstdefinestyle{javastyle}{
  language=Java,
  basicstyle=\ttfamily\small,
  keywordstyle=\bfseries\color{blue!60!black},
  commentstyle=\itshape\color{gray!70!black},
  stringstyle=\color{green!40!black},
  numbers=left,
  numberstyle=\tiny\color{gray},
  stepnumber=1,
  numbersep=8pt,
  showstringspaces=false,
  tabsize=2,
  breaklines=true,
  columns=fullflexible,
  upquote=true
}

\newcommand{\rev}[1]{\textcolor{black}{#1}}

\lstdefinestyle{json}{
    basicstyle=\ttfamily\small,
    columns=flexible,
    breaklines=true,
    frame=single,
    backgroundcolor=\color{gray!10},
    keywordstyle=\color{blue},
    stringstyle=\color{orange},
    commentstyle=\color{gray},
    morekeywords={Example_ID, RefactMethod, ZeroShotCode, InstrucCode}
}
\newcommand{\COMMENT}[1]{}

\usepackage{xspace}

\newtcolorbox{resultbox}{colback=gray, arc=0.5mm, top=1mm, bottom=1mm, left=1mm, right=1mm}

\newboolean{showcomments}

\definecolor{codegreen}{rgb}{0,0.6,0}
\definecolor{codegray}{rgb}{0.5,0.5,0.5}
\definecolor{codepurple}{rgb}{0.58,0,0.82}
\definecolor{backcolour}{rgb}{0.95,0.95,0.92}
\definecolor{javapurple}{rgb}{1,0.5,0.31} 
\definecolor{javablue}{rgb}{0.25,0.35,0.75}
\definecolor{backbefore}{rgb}{0.94,0.63,0.63}
\definecolor{backafter}{rgb}{0.6,1,0.6}

\definecolor{gray}{rgb}{0.83,0.83,0.83}
\definecolor{green}{rgb}{0.56,0.93,0.56}

\setboolean{showcomments}{true}

\ifthenelse{\boolean{showcomments}}
{\newcommand{\nb}[2]{
		\fbox{\bfseries\sffamily\scriptsize#1}
		{\sf\small$\blacktriangleright$\textit{#2}$\blacktriangleleft$}
	}
}
{\newcommand{\nb}[2]{}
}

\usepackage{listings}

\lstdefinestyle{mystyle}{backgroundcolor=\color{backcolour},commentstyle=\color{codegreen},
  keywordstyle=\color{javapurple},
  numberstyle=\tiny\color{codegray},
  stringstyle=\color{codepurple},
  basicstyle= \textnormal,
  breakatwhitespace=false,         
  breaklines=true,                 
  captionpos=b,                    
  keepspaces=true,                 
  numbers=left,                    
  numbersep=5pt,                  
  showspaces=false,                
  showstringspaces=false,
  showtabs=false,                  
  tabsize=2,
}
\begin{document}

\title{
Refactoring with LLMs: Bridging Human Expertise and Machine Understanding}


\authorrunning{Chen Kuang Piao et al.} 

\author{Yonnel Chen Kuang Piao         \and
       Jean Carlors Paul \and Leuson Da Silva \and Arghavan Moradi Dakhel \and Mohammad Hamdaqa \and Foutse Khomh
}

\institute{Yonnel Chen Kuang Piao         \and
       Jean Carlors Paul \and Leuson Da Silva \and Arghavan Moradi Dakhel \and Mohammad Hamdaqa \and Foutse Khomh \\
        Department of Computer Engineering and Software Engineering \\
        Polytechnique Montreal \\
        Montreal, QC, Canada \\
        \email{yonnel.chen-kuang-piao@etud.polymtl.ca, jean-carlors.paul@etud.polymtl.ca, leuson-mario-pedro.da-silva@etud.polymtl.ca, arghavan.moradi-dakhel@etud.polymtl.ca, mohammad-adnan.hamdaqa@polymtl.ca, foutse.khomh@polymtl.ca}
}

 \date{}

\maketitle

\begin{abstract}
Code refactoring is a fundamental software engineering practice aimed at improving code quality and maintainability. Despite its importance, developers often neglect refactoring due to the significant time, effort, and resources it requires, as well as the lack of immediate functional rewards. Although several automated refactoring tools have been proposed, they remain limited in supporting a broad spectrum of refactoring types.

In this study, we explore whether instruction strategies inspired by human best-practice guidelines can enhance the ability of Large Language Models (LLMs) to perform diverse refactoring tasks automatically. Leveraging the instruction-following and code comprehension capabilities of state-of-the-art LLMs (e.g., GPT-mini and DeepSeek-V3), we draw on Martin Fowler’s refactoring guidelines to design multiple instruction strategies that encode motivations, procedural steps, and transformation objectives for 61 well-known refactoring types.

We evaluate these strategies on benchmark examples and real-world code snippets from GitHub projects. Our results show that instruction designs grounded in Fowler’s guidelines enable LLMs to successfully perform all benchmark refactoring types and preserve program semantics in real-world settings—an essential criterion for effective refactoring. Moreover, while descriptive instructions are more interpretable to humans, our results show that rule-based instructions often lead to better performance in specific scenarios. Interestingly, allowing models to focus on the overall goal of refactoring, rather than prescribing a fixed transformation type, can yield even greater improvements in code quality.
\end{abstract}

\section{Introduction}

Software systems evolve continuously throughout their lifecycle to address reported issues and adapt to changing stakeholder requirements. 
In this context of rapid and iterative development cycles—where new features must be delivered quickly and deadlines are tight—there is a heightened risk of deploying poor or inefficient code to production \citep{Shirafuji_2023}. 
To mitigate this risk, software engineers rely on community best practices that help maintain long-term code quality and system maintainability. 
Among these practices, code refactoring has become one of the most well-known and widely adopted approaches in the software engineering community \citep{mens2004survey}.

Refactoring encompasses a wide range of transformations aimed at improving the internal structure of code without altering its external behavior. 
Martin Fowler’s seminal work \citep{Fowler2018} provides one of the most comprehensive frameworks for understanding and applying refactoring in practice. 
His catalog systematically classifies 61 distinct refactoring types, offering detailed descriptions, motivations, procedural steps, and illustrative examples to help developers understand when and how to apply each transformation effectively. 
Over time, Fowler’s catalog has become a cornerstone reference for both research and practice, guiding studies and tools in code transformation and software quality improvement \citep{Fowler_Kim2014, Fowler_Tavares2018, Fowler_Brito2020, Fowler_Rahman2022, Fowler_Niu2024, Fowler_Hasan2024}.

Refactoring typically involves several key steps: identifying problematic code segments, selecting appropriate refactoring types, applying transformations, and validating behavioral equivalence after refactoring \citep{Fowler2018}. Traditionally, this process has been performed manually by developers, guided by best practices and their own experience. 
However, despite its benefits, developers often hesitate to engage in refactoring due to its time-consuming and error-prone nature \citep{Shirafuji_2023}. 
This has motivated extensive research into automating refactoring to reduce manual effort and improve consistency. 
Early automation efforts led to the integration of refactoring tools into IDEs such as Eclipse, supporting a limited subset of refactoring types \citep{Eclipse_extract_class_2012, Eclipse_extract_method_2014, Eclipse_fuhrer_2007}. 
Yet, as Eilertsen et al. (\citeyear{Eilertsen2021}) report, developers frequently avoid these tools, citing their limited ability to handle complex or diverse refactorings.

To complement these tools, several approaches have focused on detecting refactorings automatically, such as RefactoringMiner \citep{TsantalisRMiner2018} and RefFinder \citep{Kim2010Ref-Finder}. 
While these tools provide valuable insights into refactoring activities across large-scale software repositories, they cannot identify which refactoring type should be applied to a specific code snippet, nor can they perform the refactoring autonomously.

Recent advances in Large Language Models (LLMs) have opened promising opportunities to address these limitations. 
Trained on extensive datasets of human-written code and natural language, LLMs demonstrate strong code comprehension and instruction-following capabilities. 
Consequently, they have been applied to a growing range of software engineering tasks—including automatic code refactoring \citep{Shirafuji_2023, Liu2023RefBERT, Pomian2024, Choi2024}.

For example, Liu et al. (\citeyear{Liu2023RefBERT}) proposed RefBERT, a model specialized for one refactoring type, \textit{Renaming}, to improve code readability and consistency. 
Cordeiro et al. (\citeyear{Starcoder_2024}) later explored using LLMs to refactor open-source Java projects, focusing on reducing code smells and improving code quality. 
Their results showed that, even without explicit instructions, LLMs can successfully perform simple refactorings such as \textit{Rename Method} and \textit{Extract Method} \citep{Starcoder_2024}. 
However, their findings also revealed a key limitation: LLMs tend to perform well on simpler transformations but struggle with more complex refactoring types requiring deeper reasoning about design intent and software quality. 
This gap highlights the need for more sophisticated and versatile LLM-based refactoring approaches capable of handling a broader and more challenging spectrum of refactoring types.

To address this gap, our study leverages the instruction-tuning capabilities of LLMs to enable them to perform a wider variety of refactoring tasks. 
We draw on established software engineering principles—particularly Fowler’s catalog \citep{Fowler2018}—to design instruction strategies that incorporate both the motivation and procedural steps behind each refactoring type. 
These structured instructions guide LLMs to apply diverse transformations while maintaining semantic correctness and improving code quality.

Specifically, our study focuses on three critical aspects of the refactoring process:
(i) applying refactorings,
(ii) preserving semantic behavior after refactoring, and
(iii) enhancing overall code quality.
To explore these aspects, we compare two distinct models—OpenAI’s closed-source model \gptmini\ and the open-source model DeepSeek-V3. We design multiple instruction styles enriched with contextual information from Fowler’s catalog to examine how different formulations influence model performance across 61 refactoring types.

Our investigation is guided by two research questions:

\begin{itemize}
    \item \textbf{RQ1:} How can instruction strategies inspired by best-practice guidelines guide LLMs in applying a diverse range of refactoring types? 
    
    This question evaluates whether human-oriented guidelines, when translated into machine-readable instructions, enable LLMs to apply different refactorings correctly while preserving semantics.

    \item \textbf{RQ2:} What is the impact of different instruction strategies on the quality of refactored code produced by LLMs?
    
    This question assesses how instruction design affects code quality improvements beyond correctness.
\end{itemize}

Our results demonstrate that instruction design plays a crucial role in shaping LLM performance. 
Rule-based instructions—adapted from heuristics used in automated refactoring detection—enable \gptmini\ to produce compilable, semantically preserved code more effectively than step-by-step, human-oriented instructions. 
Moreover, when LLMs are given only the high-level goal of refactoring, rather than explicit procedural guidance, they often generate code with higher overall quality (e.g., reduced complexity). 
This suggests that allowing LLMs to reason about the objective of refactoring rather than enforcing a fixed transformation can yield more meaningful improvements.

We also observe that certain localized refactorings (e.g., variable-level changes) are consistently easier for LLMs to perform across instruction types. 
Across models, DeepSeek-V3 achieved perfect success on 48 refactoring types in our benchmark dataset, whereas \gptmini\ achieved full success on 14 types.

This study makes the following contributions:

\begin{itemize}
    \item \textbf{Instruction design and evaluation:} We design and assess diverse instruction styles derived from best-practice refactoring guidelines to analyze how they influence LLMs’ ability to perform refactorings.

    \item \textbf{Enhanced automation capability:} We extend the refactoring capacity of LLMs to a broader set of refactoring types while ensuring semantic preservation and measurable code quality improvement.

    \item \textbf{Open replication package:} We release a complete replication package containing datasets, scripts, and an automated evaluation framework to assess semantic preservation and code quality after refactoring \citep{appendix}.
\end{itemize}

The remainder of this paper is structured as follows. Section~\ref{related_work} reviews related work. Section~\ref{sec:methodology} describes our methodology, and Section~\ref{sec:results} presents our findings. Section~\ref{sec:discussion} discusses the implications of our results, Section~\ref{sec:threats} outlines threats to validity, and Section~\ref{sec:conclusion} concludes the paper.

\section{Related Work} \label{related_work}

In this section, we discuss related work relevant to our study. First, we discuss state-of-the-art tools and techniques used to assist in dealing with refactorings. 
Then, we discuss how LLMs are used to perform refactoring. 

\subsection{Code Refactoring} \label{sec:refact}

Al Dellal and Abdin \citep{refact_impact_dellal} conducted a systematic literature review to analyze empirical evidence on the impact of object-oriented code refactoring on software quality attributes, synthesizing results from 76 primary studies. The review found that, while refactoring techniques generally improve internal quality attributes, such as cohesion and complexity, their impact on external quality attributes, such as flexibility and maintainability, is more variable. For example, refactorings like \textit{Extract Class} and \textit{Extract Method} were shown to reduce cyclomatic complexity by up to 30\% and improve cohesion by approximately 15\%. However, these refactoring techniques can sometimes lead to trade-offs, such as increased coupling or negative impacts on maintainability, with certain studies reporting up to a 10-15\% increase in coupling in some scenarios. Additionally, the paper highlighted that about 62\% of studies showed a positive impact on maintainability, while a smaller proportion (12\%) indicated negative effects under specific circumstances. These findings underscore the complex and context-dependent nature of refactoring, suggesting that, while it can be beneficial, its effects on software quality should be carefully assessed in each case.

\subsection{Tools for Refactoring and Code Analysis}\label{sec:tools_for_refact}

Over time, assistive tools that support practitioners in performing refactorings have proven to be highly relevant in daily programming tasks for detecting and applying refactorings.
Among these tools, COMEX \citep{DasComex2023} stands out as a framework that enables developers to generate code representations, such as Control Flow Graphs (CFG) and Abstract Syntax Trees (AST), which can be leveraged by LLMs for software engineering tasks. This approach allows LLMs to rely on structural and semantic code properties, rather than simply on an arrangement of tokens.

Another notable approach to assistive code refactoring is the use of heuristic search algorithms. In this context, the tool Opti Code Pro \citep{opticodepro} leverages this approach to guide the refactoring process, assessing potential modifications through predefined heuristics. By implementing a best-first search algorithm, the tool is guided by heuristic functions that evaluate the impact of refactorings on key software quality metrics. The results demonstrated that Opti Code Pro was effective in identifying beneficial refactoring opportunities, leading to improvements in code quality.

Furthermore, RefactoringMiner 2.0 \citep{TsantalisRefactoringMiner2.0-2022} is widely used for detecting refactorings in Git repositories. In contrast to its predecessor \citep{TsantalisRMiner2018}, the new release supports submethod-level refactorings, such as \textit{Extract Method} and \textit{Rename Method}.
In its evaluation, RefactoringMiner 2.0 reported high precision and recall rates (99.6\% and 94\%, respectively), making it one of the most reliable tools for automated refactoring detection.
Similarly, Ref-Finder \citep{Kim2010Ref-Finder} is another tool designed to detect refactorings in Java code by analyzing structural changes within the source code. Ref-Finder operates as a rule-based tool, utilizing predefined rules to identify various refactoring types. To function, the tool requires both the pre- and post-change versions of a Java repository to identify transitions that can be classified as refactorings. Compared to other tools, Ref-Finder covers a broader range of refactoring types (63 out of the 72 types outlined in the first edition of Fowler’s catalog).

Regarding tools for performing refactorings, practitioners are typically supported by IDEs, such as IntelliJ IDEA \citep{intellij}, Eclipse \citep{eclipse}, and Visual Studio Code \citep{vscode}, which provide automated refactoring services either natively or through plugins. Notable examples include the JRefactory \citep{jrefactory} and Eclim \citep{eclim} plugins for Eclipse \citep{jrefactory}, as well as the native support for simple refactoring tasks provided by Visual Studio Code \citep{vs_code_refact} and IntelliJ IDEA \citep{intellij_refact}. These services offer essential functionalities such as \textit{Extract Method}, \textit{Extract Variable}, and \textit{Rename Variable}. However, while they are effective at applying simple and moderate refactoring types, their scope remains limited to basic code changes, and they generally fall short when it comes to supporting more complex and advanced refactoring types. As a result, developers are often left with limited support for more intricate refactoring tasks.

\subsection{LLMs In Code Refactoring}\label{sec:llm_refact}

Large Language Models have revolutionized the development of software, particularly due to their ability to generate and summarize code \citep{Wang2021, ChenCodex2021, NijkampCodeGen2022}. As a result, the application of LLMs in code refactoring has gained significant attention in recent years. Although this area of research is still emerging, several studies have already explored its potential and made notable advancements \citep{Shirafuji_2023, Pomian2024, Liu2023RefBERT, Choi2024}.

Similar to our current work, Shirafuji et al. (\citeyear{Shirafuji_2023}) utilized an earlier version of OpenAI's LLM, GPT-3.5, to produce less-complex Python code by providing the model with zero-shot, one-shot, and few-shot examples aimed at reducing code complexity. As a result, they demonstrated that 95.68\% of programs could be refactored, reporting a decrease in both code complexity and lines of code for the semantically correct programs, with average reductions of 17.35\% and 25.84\%, respectively.

Pomian et al. (\citeyear{Pomian2024}) explore how to combine the limitations of LLMs with the static code analysis capabilities of IDEs to support the \textit{Extract Method} refactoring. Their preliminary study, which involved a sample of 1,752 \textit{Extract Method} scenarios using LLMs, found that up to 76.3\% of the suggestions were incorrect due to hallucinations \citep{hallucinations}. To address this issue, the authors proposed EM-Assist, an IntelliJ IDEA plugin that leverages LLMs to suggest more relevant method extractions compared to traditional IDE plugins. Their results showed that LLMs performed more successfully when contextualized by the static analysis provided by the IDE tools.

With a similar goal in mind, Liu et al. (\citeyear{Liu2023RefBERT}) proposed a two-stage framework, RefBERT, to automate the \textit{Rename} refactoring process. The framework utilizes bidirectional encoder representations from transformers (BERT), a pre-trained model originally designed for natural language processing tasks \citep{Devlin2019}.
RefBERT consists of two stages: (i) providing context using the BERT model and (ii) leveraging that context to generate a meaningful new name based on established naming conventions. The authors demonstrated that context-aware LLMs produce better results compared to scenarios where no context is provided.

While previous studies have demonstrated the efficiency of context-aware LLMs, Choi et al. (\citeyear{Choi2024}) showed how an iterative approach can further enhance LLM performance in the context of code refactoring, specifically in terms of readability and maintainability. The authors propose a solution that first identifies the method with the highest code complexity (CC) and then applies a refactoring to reduce it. This process is repeated iteratively until a satisfactory result is achieved. Additionally, to ensure that the original behavior is preserved, regression tests are conducted, and any changes that lead to test failures are rejected. The evaluation of their approach showed that the average CC could be reduced by up to 10.4\% after 20 iterations.

This study expands on previous studies by exploring the application of LLMs in code refactoring, focusing on a wider variety of refactoring types. It also introduces a comprehensive evaluation of five distinct prompt strategies, providing deeper insights into the diverse capabilities of LLMs for automating the refactoring process. By considering various approaches and refactoring scenarios, this study seeks to provide a more nuanced understanding of how LLMs can be utilized to improve code quality and maintainability.

\section{Methodology}
\label{sec:methodology}

\begin{figure}[t]
    \centering
    \includegraphics[width=\textwidth]{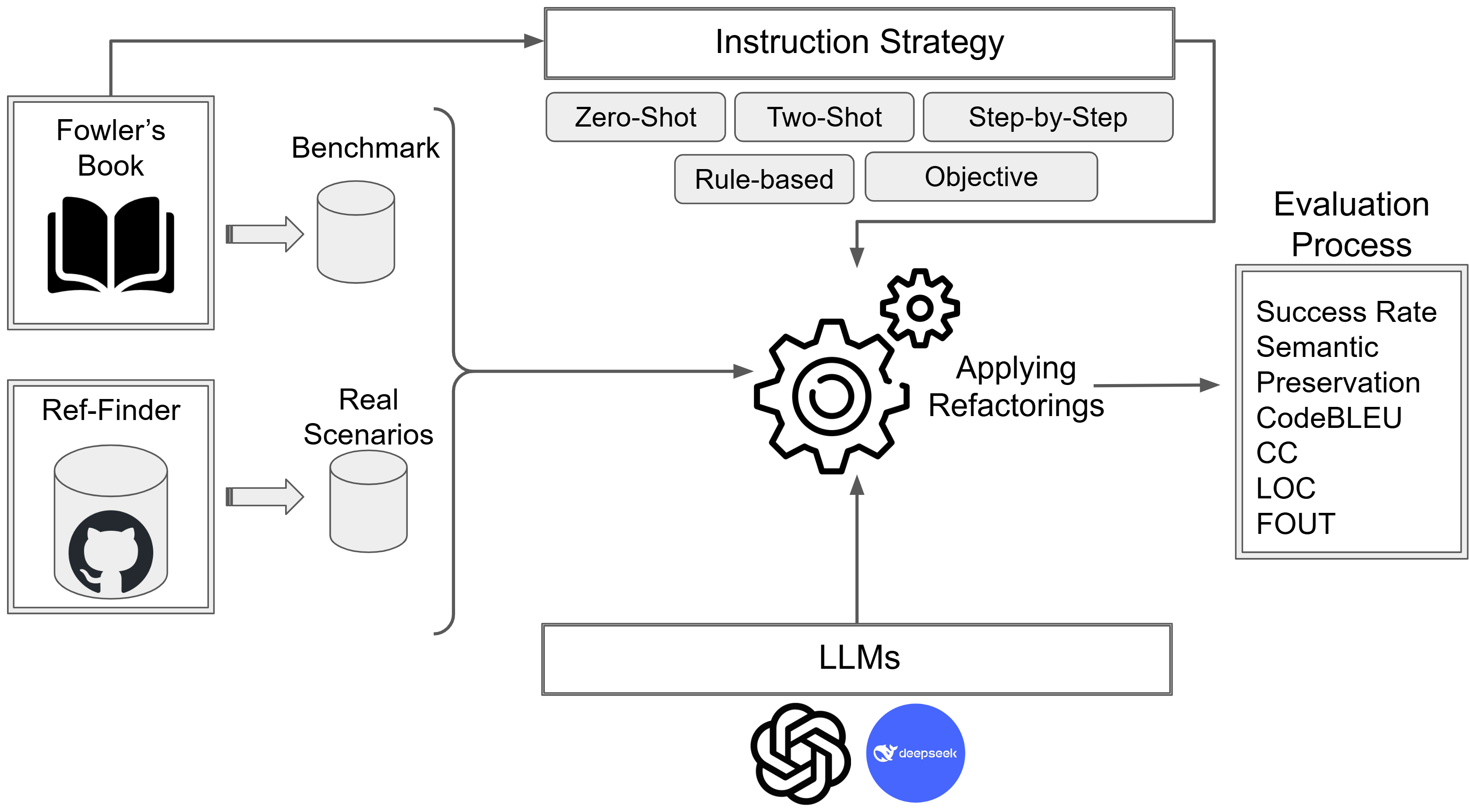}
    \caption{Empirical setup.
    }
    \label{fig:methodology}
\end{figure}

In this section, we present the methodology adopted to conduct our study (see Figure \ref{fig:methodology}).
First, we explain how we establish the datasets of benchmark and real refactoring scenarios. Next, we discuss the different instructions considered, followed by the selection of LLMs.
Finally, once we generate the refactorings for the scenarios under evaluation, we present the metrics used to guide our analysis. 

\subsection{Data Collection - Refactoring Scenarios}
\label{sec:data-collection}

In this section, we explain the process of collecting our datasets. This study considers two datasets, covering \rev{benchmark} and real scenarios of refactorings. 
In this section, we explain the process for establishing these different datasets, and how we collected the cases. 

\subsubsection{\rev{Benchmark} Scenarios - Fowler's Catalog}
\label{Fowler data collection}

\begin{figure}[h]
    \centering
    \includegraphics[width=\textwidth]{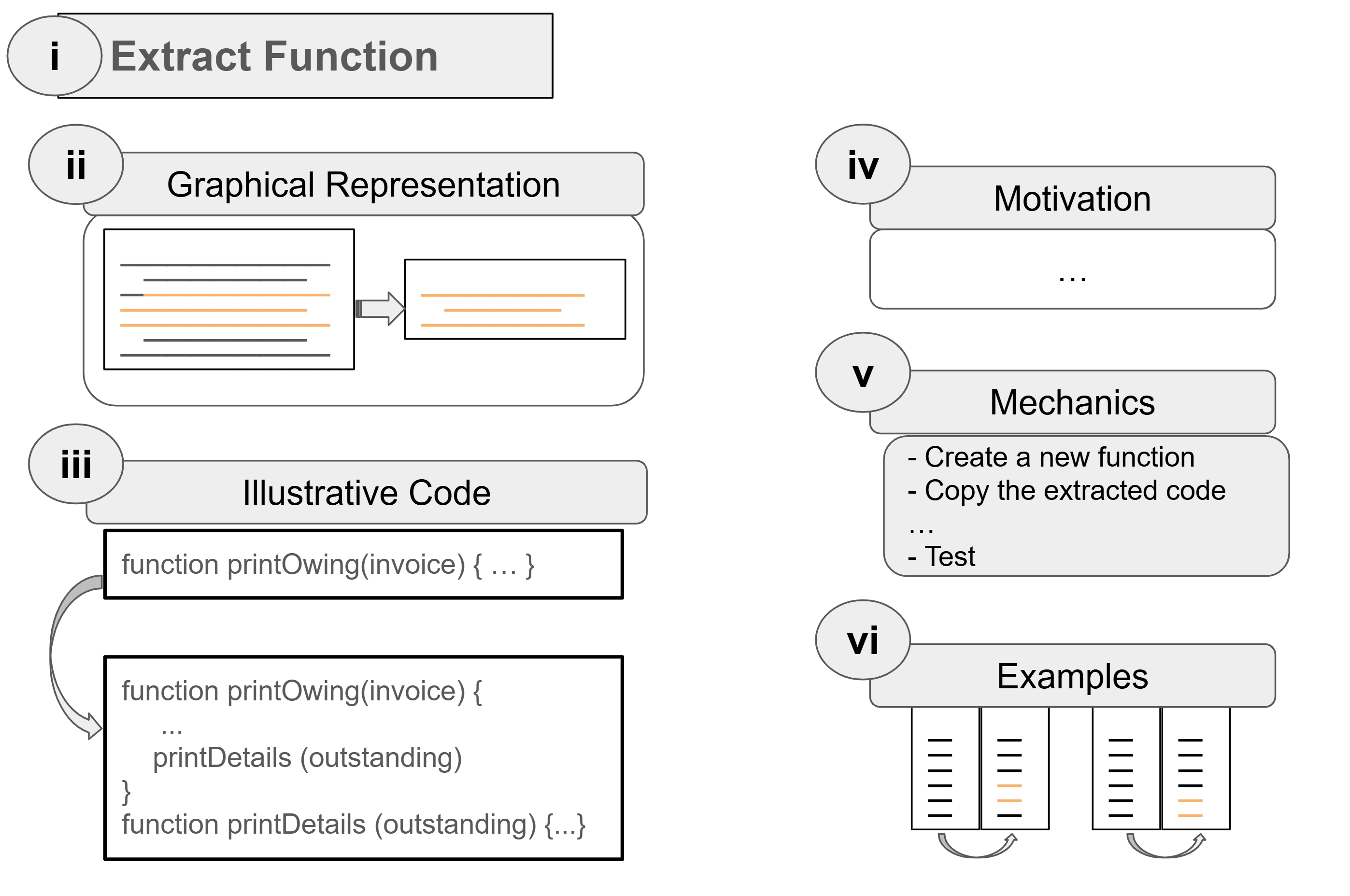}
    \caption{Extract Function: refactoring information collected from Fowler's Book.}
    \label{fig:example-refactoring}
\end{figure}

Martin Fowler's guideline aggregates a set of 61 refactoring types \citep{Fowler2018}, grouped based on their major goals and specifications.
\rev{Each refactoring is characterized by the following metadata}: (i) name, (ii) graphical representation, (iii) illustrative code snippet, (iv) motivation, (v) step-by-step instructions for applying the refactoring, and, in most cases, (vi) real-world code examples.
Figure \ref{fig:example-refactoring} presents an example of the information provided for \textit{Extract Function}.

\rev{Based on the wide range of references available for refactoring, Fowler’s catalog serves as our primary reference for two key reasons. First, it is widely used as a foundational work in the community by providing a systematic framework for identifying and implementing refactoring techniques that improve code quality and reduce maintenance costs \citep{Fowler_Kim2014, Fowler_Tavares2018, Fowler_Brito2020, Fowler_Rahman2022, Fowler_Niu2024, Fowler_Hasan2024}. Second, the book offers a comprehensive collection of refactoring techniques accompanied by practical guidelines in multiple formats, from illustrative code examples to step-by-step instructions, that make it a solid theoretical grounding for our study in designing different types of instructions for LLMs.}

We use Fowler's book in two directions: (i) adopting the illustrative code snippets of refactoring as a benchmark dataset of diverse refactoring scenarios, which we refer to as \textit{Benchmark Scenarios} and (ii) collecting the motivations, step-by-step procedures, and examples to design different instructions for our instruction-learning step (see Section \ref{sec:llm_prompting}).
For the Benchmark Scenarios dataset, we manually gathered the pre- and post-refactoring code snippets for each refactoring type from Martin Fowler's official Catalog of Refactorings (step 3 in Figure \ref{fig:example-refactoring}).\footnote{https://refactoring.com/catalog/}
The catalog provides a well-organized set of all illustrative code snippets featured in the second edition of Fowler's book. 
The Benchmark Scenarios dataset consists of all 61 refactoring types in this catalog. 

\subsubsection{Real Scenarios - 
 Ref-Finder Tool \citep{Kadar2016Dataset}}
\label{sec:real_scenario}
To gather real scenarios of refactorings aligned with Martin Fowler's catalog, we selected scenarios previously reported by Kádár et al. (\citeyear{Kadar2016Dataset}), collected from GitHub repositories using the Ref-Finder tool \citep{kim2010ref}.
\rev{Ref-Finder supports 61 refactoring types from Fowler's catalog, and it is a well-known tool with a good overall accuracy in detecting those refactorings.
Furthermore, the refactoring scenarios reported in this study were manually validated by the authors after being identified through Ref-Finder.} In their study, Kádár et al. (\citeyear{Kadar2016Dataset}) built their dataset using seven open-source Java projects. 
The dataset comprises 7,872 samples covering 19 method-level refactoring types, of which 626 were manually validated by the authors. Their manual labeling process resulted in 145 True Positive (TP) cases. 

To ensure the quality of our dataset, we first relied on 145 TP samples from the validated dataset of Kadar's study. 
Out of this sample-set, we were able to select 35 scenarios, distributed across 5 method-level refactoring types, for which we could compile the pre- and post-refactoring versions (commits) of the repositories.
During this process, we prioritized projects that we were able to compile their pre- and post-refactoring versions, starting with ANTLR4\footnote{\href{https://github.com/antlr/antlr4}{https://github.com/antlr/antlr4}}, followed by JUnit\footnote{\href{https://github.com/junit-team/junit4}{https://github.com/junit-team/junit4}}. 
For the remaining TP examples collected from other projects, we were unable to compile and execute the version of the project that was used to collect the examples.

To expand our dataset, we manually validated additional examples from the original dataset collected from the two projects of ANTLR4 and JUnit that had not previously been manually validated. Ultimately, our dataset encompassed 53 real scenarios across 11 distinct refactoring types for which we were able to compile both pre- and post-refactoring code examples. Some refactorings, such as \textit{Extract Method}, contained a larger number of validated examples, whereas others had only a few, and certain types—such as \textit{Replace Subclass with Delegate}—yielded no valid examples. We refer to this dataset as \textit{Real Scenarios}.

\subsubsection*{Manual Validation Process}
We manually validated additional examples that had not previously undergone validation to expand our dataset with more real scenarios. To do so, we followed the same methodology adopted in the original study to conduct the manual labeling process \citep{Kadar2016Dataset}. 
For each example, two researchers independently evaluated the instance and its corresponding refactoring type as labeled by Ref-Finder. Disagreements were resolved through discussion (5 out of 42 cases). Ultimately, the dataset includes only those samples for which both validators agreed with the label assigned by Ref-Finder and for which both pre- and post-refactoring versions of the code could be compiled, resulting in 18 additional validated cases.

\vspace{0.5em}
\rev{
It is worth noting that the main challenge in constructing a refactoring dataset that also enables assessment of semantic preservation across diverse refactoring scenarios lies in the ability to compile and execute both the pre- and post-refactoring versions of the code examples. This challenge is further amplified by the high dependency management requirements of Java projects. As part of our replication package, we release the dataset with executable real-world scenarios and organize a framework to automatically conduct the assessment (see Section~\ref{sec:data-processing}). Each refactoring is represented by a unique key, with its type and corresponding code provided as values, as shown in Listing \ref{listing_realistic}. }

\begin{lstlisting}[style=json, caption={Real scenario Example 01: Extract Variable}, label=listing_realistic]
{
  "REALISTIC_Example_01": {
    "RefactMethod": "EXTRACT VARIABLE",
    "BeforeCode": "function printOwing(invoice) {...}",
    "commitID_before": "ad9bac95",
    "path_before": "...ParserFactory.java",
    "name": "List<SrcOp> set(...)",
  }
}
\end{lstlisting}

\subsection{Instruction Learning}
\label{sec:llm_prompting}
Once the refactoring datasets were collected, we proceeded to design the \rev{instructions} used when calling the LLMs.
As previously shown in Figure \ref{fig:example-refactoring}, Fowler’s book provides an overview of various refactorings that can serve as guidelines for developers to conduct those refactorings in practice. We rely on the different information provided for each refactoring type to design a variety of instructions. Based on this information, we parsed the PDF version of the book and extracted the following elements to construct the instructions: (i) the name of the refactoring, (ii) the step-by-step instructions for applying it, and (iii) code examples (steps 1, 4, and 6, respectively, in Figure \ref{fig:example-refactoring}).

In the following, we explain how the extracted information was used to construct the different instructions employed in this study. 
\rev{Overall, these different instructions aimed at investigating whether the level of information would play a role in how LLMs understand and apply refactoring task. 
Each strategy differs in how information about the refactoring types is presented and the level of detail.}

\subsubsection{Zero-Shot Learning}

Zero-Shot learning is a well-known technique in which a model is prompted to perform a task \citep{Wang2019ZSL} without additional instruction. \rev{Given that LLMs are trained on diverse data available on the internet, possibly including Fowler’s catalog, we use this instruction type—where only the name of the refactoring is provided— to assess whether the models already possess knowledge of well-known refactorings and can apply a specific refactoring without any additional instruction or explanation.}
Following this approach, we constructed our prompt by asking the LLMs to apply a given refactoring type to a code snippet without providing any additional context. For example, in the case of the \textit{Extract Variable} refactoring, the prompt includes only the name of the refactoring. The following template is used, where \texttt{refactoring\_name} represents the desired refactoring type and \texttt{code} refers to the code segment to be refactored.

\begin{tcolorbox}[myinstructionbox, title=Zero-Shot Learning]
\textit{Apply the \texttt{\$<refactoring\_name>\$} refactoring on the following java code: \texttt{\$<code>\$}}

\vspace{0.5em}

\textit{Generate the final code in java. Clean the output to only show the final version of the code and do not include non-programming language content.}
\end{tcolorbox}

\subsubsection{Two-Shot Learning}

Few-shot learning occurs when the model is provided with a small set of supervised examples—two in this case—to enhance its performance \citep{Wang2020FSL}. As discussed in previous sections, Fowler’s catalog provides code examples for each refactoring type to guide developers in applying them. \rev{Each code example presents the original version of the code, followed by a sequence of modifications made during the refactoring process, and ends with the final refactored version of the code.} The two-shot learning instruction is designed to evaluate the performance of LLMs when given such examples as additional context. With this instruction, we aim to investigate whether providing supervised code examples guides LLMs to generate more accurate refactorings. For this purpose, we provide the LLM with two code examples from Fowler’s catalog. In cases where two code examples are not available for a given refactoring type at step 6 in Figure~\ref{fig:example-refactoring}, we use an illustrative snippet from step 2 in Figure~\ref{fig:example-refactoring} and exclude that snippet from the \textit{Benchmark Scenarios} dataset.
\rev{Note that we do not report separate results for the \textit{Two-Shot} learning on the \textit{Benchmark Scenarios}, since these examples are directly derived from Fowler’s catalog and already form the basis of this dataset, which could bias the LLM's behavior. We applied this instruction only on \textit{Real Scenarios} dataset.}
 
The following template is used for this instruction, where \texttt{refactoring\_name} represents the desired refactoring type, \texttt{refactoring\_examples} refers to the two-shot examples, and finally, \texttt{code} refers to the code segment to be refactored.

\begin{tcolorbox}[myinstructionbox, title=Two-Shot Learning]
\textit{Following are examples to apply the \texttt{\$<refactoring\_name>\$} refactoring: \texttt{\$<refactoring\_examples>\$}}

\vspace{0.5em}

\textit{Now, given the following code, apply the \texttt{\$<refactoring\_name>\$} refactoring: \texttt{\$<code>\$}}

\vspace{0.5em}

\textit{Generate the final code in java. Clean the output to only show the final version of the code and do not include non-programming language.}
\end{tcolorbox}

\subsubsection{Step-by-Step Learning}
\label{sec:step_by_step_learning}
\rev{Different from the previous instructions, for the \textit{Step-by-Step} instruction, we take a different route by providing the LLMs with a step-by-step guide collected from Fowler's catalog on how to perform the refactoring, as previously described and shown in Figure \ref{fig:example-refactoring} (step 5).} 
This approach is tailored to each refactoring type and defines the specific steps required to perform it. The instruction in this prompt is intended to evaluate the performance of LLMs when provided with explicit step-by-step guidelines, similar to those originally designed to assist developers in implementing refactorings.
For instance, the steps for the \textit{Extract Variable} refactoring are as follows \citep{Fowler2018}:
\begin{itemize}
    \item Ensure that the expression you want to extract does not have side effects;
    \item Declare an immutable variable. Set it to a copy of the expression you want to name;
    \item Replace the original expression with the new variable.
\end{itemize}

For this instruction, the following prompt template is used, where \texttt{refactoring\_name} represents the desired refactoring type, \texttt{steps} represents the set of instructions to perform the desired refactoring type, and \texttt{code} refers to the code snippet to be refactored.

\begin{tcolorbox}[myinstructionbox, title=Step-by-Step Learning]
\textit{Following are step-by-step instructions on how to apply the \texttt{\$<refactoring\_name>\$} refactoring:
\texttt{\$<steps>\$}}

\vspace{0.5em}

\textit{Now, given the following code, apply the \texttt{\$<refactoring\_name>\$} refactoring on it: \texttt{\$<code>\$}}

\vspace{0.5em}

\textit{Generate the final code in java. Clean the output to only show the final version of the code and do not include non-programming language.}
\end{tcolorbox}

\subsubsection{Rule-based Learning}
\label{sec:rule-based-learning}
\rev{The Ref-Finder tool~\citep{Kadar2016Dataset}, which is also used in this study to construct the \textit{Real Scenarios} dataset, supports a range of refactoring types from Fowler’s catalog, as discussed in Section \ref{sec:real_scenario}. In this tool, to represent each refactoring type and automatically detect its occurrence in Java code, a set of rules has been developed for each type. These rules are embedded within the tool to enable automated detection of refactorings. These rules are introduced by Prete et al.(\citeyear{rules}) and organized into a catalog \citep{rules_catalog} that describes refactoring patterns in a rule-based format, which can serve as a method for formally representing each refactoring type.}

Here is an example of the rules associated with \textit{Rename Method} (referred to as \textit{Change Function Declaration} in the second edition of Fowler's book).
The associated rule states that a valid refactoring is observed when a method is added, while another one is removed, and both methods share similar content. 
Below, the rule is presented:

\begin{tcolorbox}[myinstructionbox, title=Rule for Rename Method]
\texttt{added\_method(newmFullName, newmShortName, tFullName)}
$\land$ \\
\texttt{deleted\_method(mFullName, mShortName, tFullName)} $\land$ \\
\texttt{similarbody(newmFullName, newmBody, mFullname, mBody)} $\rightarrow$ \\
\texttt{rename\_method(mFullName, newmFullName, tFullName)}
\end{tcolorbox}

\rev{By providing this rule-based instruction, we aim to investigate whether the rules originally designed to develop an automatic refactoring detection tool—and to guide the tool in identifying different refactorings—can also guide LLMs in performing them. The key difference between this rule-based instruction and the step-by-step instruction (Section~\ref{sec:step_by_step_learning}) is that the latter consists of natural-language guidelines that are easier for humans to interpret, whereas the former relies on rule-like, machine-oriented instructions that are generally more difficult for humans to be used as guideline.}

Finally, for each supported refactoring type, the LLM is provided with (i) the refactoring type, (ii) its associated rule, and (iii) the target code to be refactored.
For this prompt, the following template is used, where \texttt{refactoring\_name} represents the desired refactoring type, \texttt{rule} represents the associated rule to perform the desired refactoring type, and \texttt{code} refers to the code segment to be refactored.

\begin{tcolorbox}[myinstructionbox, title=Rule-based Learning]
\textit{Following is the rule to apply the \texttt{\$<refactoring\_name>\$} refactoring:
\texttt{\$<rule>\$}}

\vspace{0.5em}

\textit{Now, given the following code, apply the \texttt{\$<refactoring\_name>\$} refactoring on it: \texttt{\$<code>\$}}

\vspace{0.5em}

\textit{Generate the final code in java. Clean the output to only show the final version of the code and do not include non-programming language.}
\end{tcolorbox}

While the rules by~\cite{rules_catalog} were based on the first edition of Fowler's Catalog of Refactorings, our study relies on the second edition, which could result in mismatches when tracking the rules and associated refactoring types. However, after manually reviewing each rule, we found that 46 of them could be matched to our database of 61 refactoring types, leaving 15 refactoring types without an associated rule and not considered for this type of instruction.

\subsubsection{Objective Learning}
\rev{In all the instructions discussed so far, the target refactoring type was explicitly included in the prompt template, referred to as \texttt{refactoring\_name}. In this final instruction, however, no information about the target refactoring type or required transformation is provided to the LLM.}
Instead, the LLM is given a general objective of code refactoring, \rev{rephrased from Fowler's catalog \citep{Fowler2018},} stating that the process of refactoring aims to improve readability, maintainability, and quality without altering the initial code's external behavior. The motivation behind this instruction is to evaluate whether LLMs can learn the objective of code refactoring and correctly apply the appropriate transformations without being explicitly instructed to do so.

The following template is used, where \texttt{code} refers to the code snippet to be refactored. In this instruction, the prompt template is the same for all code snippets across both datasets, with \texttt{code} being the only variable, as shown below:

\begin{tcolorbox}[myinstructionbox, title=Objective Learning]
\textit{Code refactoring is the process of changing source code for better readability, maintainability, and quality without changing its external behavior.}

\vspace{0.5em}

\textit{Given the following code, output a refactored version of it in Java: \texttt{\$<code>\$}}

\vspace{0.5em}

\textit{Clean the output to only show the final version of the code and do not include non-programming language.}
\end{tcolorbox}

\subsection{LLMs Selection}
\label{sec:llm-setup}

To evaluate the capability of LLMs to perform refactorings, we selected two models, as described here.
The first model is OpenAI’s \gptmini\ \citep{gpt4}, a state-of-the-art LLM designed to perform a wide variety of natural language processing tasks.
The second LLM selected for this study is DeepSeek's DeepSeek-V3 \citep{guo2024deepseek}, an state-of-the-art open source model that is released after \gptmini~ and widely used in recent studies
\citep{gheyi2025evaluating, depalma2024exploring,liu2025exploring}.

To prompt these models, our scripts rely on the API services they provide.
We generate access keys for each model, allowing us to interact with them through their APIs.
Regarding hyperparameters, we used the default settings for each model, including temperature and the maximum token window.

\subsection{Prompting LLMs for Applying Refactorings}
\label{sec:data-processing}

Once we had collected all the necessary information regarding the dataset and instructions of our study, we proceeded with prompting the LLMs to address the refactoring scenarios under analysis.
For each scenario and each instruction, we prompt each LLM for five runs.
To illustrate this process, consider the \rev{input (Listing \ref{listing_realistic}, previously presented in Section \ref{sec:data-collection}).}
\rev{After collecting the refactoring's information (type and code) from input, we further retrieve the required information associated with each instruction \footnote{The information used to construct the instructions is available as part of the the input JSON file in our replication package.}.}
For example, consider Listing \ref{input_each_refactoring}, which presents the definitions of each refactoring and associated examples mined from Fowler's book.
The refactoring types are used as \texttt{keys}.

\begin{lstlisting}[style=json, caption={Definition and Examples for Each Refactoring Type}, label=input_each_refactoring]
  {
  "Extract Function": {
    "Mechanics": "Create a new function, and name it...",
    "Example: Using Local Variables": "The easiest case with local variables..."
  },
  "Change Function Declaration": {
    "Mechanics": "aka: Rename Function...",
    "Example: Renaming a Function (Simple Mechanics)": "Consider this function..."
  }
}
\end{lstlisting}

To standardize the output of the LLMs and to be able to automatically incorporate their output in the next steps, we instructed them to produce results in a specific format. Despite explicitly asking the models to adhere strictly to this format (e.g., returning only the code after applying the refactoring), some post-processing was still necessary before passing the generated outputs to the next steps to conduct further assessments.
In practice, the LLMs typically return their output as a single string, which may include multiple methods and/or classes based on the refactoring type. To enable testing for compilation and semantic preservation, this output must be split accordingly.
Through manual inspection of several cases, we identified recurring patterns that enabled us to design regular expressions (RegEx) to extract the relevant code segments accurately. As a result, the LLM outputs were transformed into the following format (Listing \ref{output_llm}):

\begin{lstlisting}[style=json, caption={Formatted Output of LLMs}, label=output_llm]
  {
  "Example_ID": {
    "RefactMethod": "EXTRACT VARIABLE",
    "ZeroShotCode": {
      "methods": ["method1", "method2"],
      "classes": ["class1"],
      "others": ["other_text"]
    },
    "InstrucCode": {...}, ...
  }
}
\end{lstlisting}

\subsection{Validation}

In this section, we describe the process used to assess whether the refactorings generated by LLMs are correctly applied. Our validation process considers multiple aspects across two different datasets, taking into account their specific characteristics.

\subsubsection{Manual Validation}
\label{sec:manual-validation}

\rev{To evaluate the correctness of the code examples in the \textit{Benchmark Scenario} dataset after applying the refactorings, we performed a manual analysis, as no predefined test cases are available to automatically execute these examples and assess their correctness and semantic preservation after refactoring.} 

This manual validation involved two researchers, who independently analyzed all the refactoring scenarios from a randomly selected run and assessed their success.
For each selected run, LLMs generated refactoring for all cases, resulting in 229 cases for each LLM under analysis (resulting in 458 analyzed cases across two LLMs). 
Each analysis was performed individually, and later, they discussed any conflicts, reaching a final conclusion.
To measure the agreement between them, we calculated the Cohen's Kappa coefficient \citep{cohen}.
This process was carried out for both LLMs under evaluation.
For \gptmini\ refactorings, we observed a Cohen's Kappa coefficient of 0.864, while for DeepSeek, the coefficient was 0.774.
Overall, these high scores demonstrate a substantial agreement between the researchers, indicating that their independent evaluations were consistent.

\subsubsection{Automatic Validation}
\label{sec:automatic-validation}
\rev{For the dataset containing real scenarios, we extend our analysis to automatically assess both the syntactic and semantic correctness of the refactorings generated by LLMs. Since this dataset is derived from real GitHub projects, corresponding test suites are available for each project and can be used to evaluate the refactored code. One of the main challenges in this kind of evaluation lies in automatically integrating the LLM-generated output into the relevant Java files, recompiling the updated files, and subsequently executing the test suites. Because our study involves multiple attempts with different instructions across various refactoring scenarios, this process must be automated to ensure both efficiency and correctness.}

\rev{In this section, we describe our methodology for automatically applying the LLM-generated refactorings to the java projects. The process involves parsing the relevant Java file and generating its Abstract Syntax Tree (AST), then identifying the nodes affected by the refactoring, applying the necessary changes to reflect the refactoring, and creating a new executable branch of the project for evaluation. This solution has been packaged as a JAR file and released as part of our replication package to facilitate future studies on code refactoring \citep{appendix}. 
The entire process can be divided into three main steps, which we outline in detail below.}

\subsubsection*{Version Identification}
\rev{This step establishes the correct project version (commit) that contains the refactoring scenario under analysis. Each refactoring example in the Real Scenario dataset belongs to a specific version of its project. To ensure the reproducibility of our results, we forked the original GitHub repositories associated with the refactoring scenarios in this dataset. The process then automatically clones these forks locally and, using the metadata collected for each refactoring scenario (as described in Section~\ref{sec:data-collection}), checks out the commit corresponding to the refactoring scenario.}
\rev{Next, the scripts set up the environment to get the information required for the next steps.\footnote{The \texttt{launcher.py} file orchestrates this process.}
First, it loads the file with the LLM-generated code (e.g., Listing \ref{output_llm}) and, for each scenario, a new local path is created to save the output results.}  
It is important to note that in this step, we ensure the original commits (\texttt{before\_refactoring}) are compilable. 
This way, if the code is no longer compilable after applying the changes, we can attribute the compilation issues to the new changes.

\subsubsection*{Applying LLM-generated Refactorings}
\rev{This step begins by locating the Java files associated with the code snippets before applying the refactoring, referred to as \texttt{path\_before\_refactoring}. The file is then parsed using JavaParser\footnote{https://javaparser.org/}
to generate its AST, which is subsequently modified to reflect the refactoring. For refactorings that occur at the method level (either the entire method or a statement within a method), the process uses the associated method name (\texttt{method\_name} node) to locate its corresponding node in the AST. The method is then replaced with the new refactored version generated by the LLM. Notably, we instruct the LLM to generate the complete method that incorporates the refactoring for the method-level ones. In cases where the LLM generated additional methods—such as in \textit{Extract Method}—the process inserts the new method as a new node in the AST.}
For refactorings at the class level, the same procedure is followed, but applied to the entire class. 
 
To facilitate reproducibility and practical use, we provide the entire solution as a JAR file.
\rev{The jar file is called via a subprocess, providing the required information as arguments (see Appendix \citeyear{appendix}). 
Once the process is completed, we examine its return value to determine whether it was successful. Several states are possible as the outcome of this step. A \textit{failed attempt} occurs when the final source code cannot be parsed after integrating the LLM-generated refactoring. In such cases, we log the error for further analysis and classify the scenario as a \textit{failed attempt}. If the parsing succeeds, the process is considered a \textit{successful attempt}, then all changes are saved and committed to a new branch, enabling us to manage the different refactorings separately for further assessment.}

\subsubsection*{Validating Refactorings}
After applying and saving all the required changes, we proceed to assess the syntactic and semantic impact of the LLM-generated refactorings. This step attempts to compile \rev{the new branch after refactoring}. As mentioned in \textit{Version Identification} step, we ensured the original commits are compilable. 
Thus, if the code in the new branch is no longer compilable, it is attributed to the new changes.

\rev{For the \textit{failed attempts} discussed in previous step, the validation step further evaluates the reported logs, aiming to identify the failure causes.}
If the code is still compilable, the validation step then moves to check the semantic preservation by running the project's test suites on the new branch after applying the refactoring.
To account for possible flakiness, we repeat this process five times.
\rev{Finally, we report any test cases that result in failures or errors. To establish a baseline, we also repeat the test execution on the commit \textit{before-refactoring} and collect the testing results, since some test cases in the test suites may already fail or raise errors prior to applying any changes introduced by the refactoring.}

\subsubsection{Evaluation Metrics}
\label{sec:metrics}

We select a set of metrics that address both general aspects of source code and those specifically related to the impact of refactoring applied by LLMs on the original code.
\rev{We focus on metrics that are critical for assessing the impact of refactoring, including code quality, size, complexity, and fault proneness. 
Furthermore, the selected metrics have been commonly used in the literature to assess the effectiveness of refactoring, supporting our choices here \citep{Wang2021, Shirafuji_2023, Starcoder_2024, metrics}.}
Below, we provide further details about the selected metrics.

\textbf{CodeBLEU:} Regarding the code generated by LLMs, CodeBLEU \citep{Ren2020CodeBLEU, codepad} is an evaluation metric designed to assess the quality of model-generated code based on its similarity to a ground truth. 
This metric is an extension of the original BLEU \citep{Papineni2002BLEU}, which is more suited for natural language. 
It adds weighted n-gram matching, AST similarity, and semantic data-flow similarity to better evaluate aspects of programming languages. 
Thus, a high CodeBLEU score indicates a high similarity in both syntax and semantics when compared to a ground truth or reference code.

\textbf{LOC, CC, and FOUT:} In addition, we compute other metrics to assess the quality of the code after refactoring compared to the ground truth refactored code, such as the number of lines of code (LOC), cyclomatic complexity (CC), and the number of method calls (FOUT).\footnote{FOUT stands for Fan-Out}
While LOC represents the total number of lines in the source code, FOUT indicates the number of times a method is invoked.
CC is a metric proposed by McCabe to assess and quantify the complexity of a given source code. 
It can be calculated by analyzing the control-flow graph (CFG) of the code \citep{McCabe1976CC}.\footnote{These metrics were automatically collected by our scripts, using two Python libraries: pyccmetrics \citep{pyccmetrics} and codebleu \citep{Ren2020CodeBLEU}}.

\textbf{Success Rate: }
\rev{To assess the correctness of the refactorings applied to the \textit{Benchmark Scenarios}, we rely on human judgment, since it is not possible to execute tests on the provided code snippets (see Section~\ref{sec:manual-validation}). As discussed earlier in Section~\ref{sec:manual-validation}, one random run from each LLM under investigation was chosen and manually evaluated by two researchers. 
Each refactoring was then classified as either \textit{successful} or \textit{unsuccessful}. Next, we calculate the success rate, using the following equation:
}
\[
\text{Success rate (\%)} = \dfrac{\text{Number of successful refactorings}}{\text{Total number of attempts}}
\]

\textbf{Compilation, New Test Failed, and New Test Error:} 
\rev{For the real refactoring scenarios (see Section~\ref{sec:automatic-validation}), we measure the number of refactoring attempts that result in (i) compilable code, (ii) additional test failures, or (iii) additional test errors. 
\textit{Compilation} indicates whether the code compiles successfully after the refactoring. 
\textit{New Test Failed} captures the number of additional test failures introduced by refactoring, calculated as the difference in failing tests before and after refactoring. 
\textit{New Test Error} follows the same logic, but for test errors instead of failures.}

When presenting our metrics, we report them based on the average across the five runs, along with the standard deviation.
\rev{To assess the significance of the results associated with different LLMs attempts, we adopted statistical testing.
Since our data consists of independent (unpaired) samples and the groups have different sizes due to some attempts failing, we chose the Mann–Whitney U test, which handles unequal sample sizes and does not assume normality.
In our analysis, we adopted a significance level of $\alpha = 0.05$, which is commonly accepted in statistical tests to denote significance. 
}

\section{Results}
\label{sec:results}
In this section, we address the research questions outlined in this study.
First, we examine whether instructions inspired by best-practices that are designed to guide humans in refactoring tasks can also guide LLMs to perform a broader range of refactoring types (RQ1).
Second, we investigate to what extent the refactorings performed by LLMs improve the quality of the resulting code (RQ2).

\subsection{RQ1: How different instruction strategies inspired by best-practice guidelines can guide LLMs in applying a diverse range of refactoring types?}

\subsubsection{Benchmark Scenarios}

\begin{table}[]
\centering
\caption{Correctness by prompt type across LLM settings based on manual validation.}
\begin{tabular}{|c|l|c|c|}
\hline
\textbf{\begin{tabular}[c]{@{}c@{}}LLM \\ Setting\end{tabular}} & \textbf{Instruction Strategy} & \multicolumn{1}{l|}{\textbf{\begin{tabular}[c]{@{}l@{}}Total\\ Attempts\end{tabular}}} & \multicolumn{1}{l|}{\textbf{\begin{tabular}[c]{@{}l@{}}Success \\ Rate (\%)\end{tabular}}} \\ \hline

\multirow{4}{*}{\gptmini}                                            
    & Zero-Shot Learning       & 61 & 54.1 \\ \cline{2-4} 
    & Step-by-Step Learning    & 61 & 83.6 \\ \cline{2-4} 
    & Rules-based Learning     & 46 & 67.4 \\ \cline{2-4} 
    & Objective Learning       & 61 & 29.5 \\ \hline

\multirow{4}{*}{DeepSeek}                                       
    & Zero-Shot Learning       & 61 & 90.2 \\ \cline{2-4} 
    & Step-by-Step Learning    & 61 & 100.0 \\ \cline{2-4} 
    & Rule-based Learning      & 46 & 100.0 \\ \cline{2-4} 
    & Objective Learning       & 61 & 36.1 \\ \hline

\end{tabular}
\label{tab:bechmark-prompt-results}
\end{table}

\begin{table}[htbp]
\centering
\small
\setlength{\tabcolsep}{2pt}
\renewcommand{\arraystretch}{0.90}

\begin{tabular}{|l|c|p{0.75\textwidth}|}
\hline
\multicolumn{1}{|c|}{\textbf{\begin{tabular}[c]{@{}c@{}}LLM \\ Setting\end{tabular}}} & 
\multicolumn{1}{c|}{\textbf{\begin{tabular}[c]{@{}c@{}}Success \\ Rate (\%)\end{tabular}}} & 
\multicolumn{1}{c|}{\textbf{Refactoring Types}} \\ \hline
\multirow{20}{*}{\gptmini} &  \raisebox{-3\height}{100} & combine functions into class, consolidate conditional expression, decompose conditional, encapsulate record, extract variable, pull up field, pull up method, remove dead code, rename variable, replace inline code with function call, replace nested conditional with guard clauses, replace parameter with query, split variable, substitute algorithm \\ \cline{2-3} 
& \raisebox{-3\height}{75} & collapse hierarchy, extract class, extract function, extract superclass, inline function, inline variable, introduce parameter object, parameterize function, push down field, remove flag argument, replace conditional with polymorphism, replace constructor with factory function, replace superclass with delegate \\ \cline{2-3} 
& \raisebox{-.8\height}{66.7} & move statements into function, move statements to callers, rename field, replace loop with pipeline, replace subclass with delegate, split loop \\ \cline{2-3} 
& \raisebox{-2\height}{50} & change reference to value, encapsulate collection, encapsulate variable, hide delegate, inline class, introduce special case, pull up constructor body, remove setting method, replace primitive with object, replace type code with subclasses, separate query from modifier \\ \cline{2-3} 
& 33 & replace derived variable with query, split phase \\ \cline{2-3} 
& \raisebox{-.8\height}{25} & change function declaration, introduce assertion, remove subclass, replace function with command, replace temp with query, slide elements \\ \cline{2-3} 
& \raisebox{-2\height}{0} & change value to reference, combine functions into transform, move field, move function, preserve whole object, push down method, remove middle man, replace command with function, replace query with parameter \\ \hline
\multirow{17}{*}{DeepSeek} & \raisebox{-8\height}{100} & change function declaration, combine functions into class, consolidate conditional expression, decompose conditional, encapsulate record, encapsulate variable, extract class, extract function, extract variable, hide delegate, inline function, inline variable, introduce special case, move field, move function, parameterize function, preserve whole object, pull up constructor body, pull up field, pull up method, push down field, remove flag argument, remove middle man, remove setting method, remove subclass, rename field, replace conditional with polymorphism, replace constructor with factory function, replace function with command, replace nested condition with guard clauses, replace parameter with query, replace superclass with delegate, replace temp with query, separate query from modifier, slide elements, split variable, substitute algorithm, split phase, split loop, replace query with parameter, replace inline code with function call, replace derived variable with query, replace command with function, rename variable, remove dead code, move statements to callers, combine functions into transform, replace subclass with delegate\\ \cline{2-3} 
& \raisebox{-2\height}{75} & change value to reference, collapse hierarchy, encapsulate collection, introduce assertion, introduce parameter object, push down method, replace primitive with object, replace type code with subclasses\\ \cline{2-3} 
& 66 & replace loop with pipeline, move statements into function\\ \cline{2-3} 
& 50 & change reference to value, extract superclass\\ \cline{2-3}
& 25 & inline class \\ \hline
\end{tabular}
\caption{Success Rate by Refactoring Types }
\label{tab:bechmark-llm-results}
\end{table}

Table \ref{tab:bechmark-prompt-results} reports the success rate of the evaluated LLMs in applying refactorings, 
covering the 61 refactoring types based on Fowler's catalog collected using different instruction strategies (\textit{Zero-Shot}, \textit{Step-by-Step}, \textit{Rule-based}, and \textit{Objective-based}). 
\rev{As explained in Section \ref{sec:metrics}, the success rate is computed based on human analysis, which serves as the ground truth in this evaluation.}
\rev{Overall, we observe that instruction strategy plays an essential role in the effectiveness of LLMs when applying refactorings. 
In particular, \textit{Rule-based} and \textit{Step-by-Step} instructions yield the highest success rates across both LLMs, with DeepSeek achieving an 100\% success rate with both instructions. 
Meanwhile, \gptmini\ reports its highest rate with the \textit{Step-by-Step} Learning, showing that such instructions might leverage the model’s ability to break down the transformation into smaller reasoning steps.
This suggests that guiding the model through structured intermediate reasoning, or by constraining it with explicit refactoring rules, can improve the accuracy of the applied transformations compared to less constrained strategies.}

\rev{By contrast, the \textit{Objective-based} prompt produces the lowest performance for both LLMs (29.5\% and 36.1\% for \gptmini\ and DeepSeek, respectively).
Such a prompt does not consider any guidance about the target refactoring, which leads the LLMs to present poor performance, mainly due to their limited ability to infer the precise transformation exclusively from a high-level objective description.
By adding the name of refactoring through the \textit{ZeroShot} prompt, we observe better results for both models. 
However, we observe that without explicit rules or step-wise instructions, the models tend to generate either incomplete or incorrect refactorings, highlighting the importance of structured prompting to align model behavior with the desired refactoring goals.
These results indicate that instructions providing explicit guidance on how to perform refactorings improve model performance. 
Nevertheless, although LLMs can recognize the general goal of refactorings and their various types, they still struggle to consistently translate this understanding into correct applications without further guidance (e.g, Objective-based Learning).}

\rev{Table \ref{tab:bechmark-llm-results} presents the success rate for each LLM across the refactoring types.
Consistent with our previous findings, DeepSeek outperforms \gptmini, achieving a perfect success rate (100\%) on 48 refactoring types, compared to only 14 for \gptmini, while also covering all cases supported by \gptmini.
Moreover, DeepSeek reports a success rate below 50\% for only a single refactoring, whereas \gptmini\ falls below this threshold for 17 cases. DeepSeek maintains higher success rates across the full distribution of refactorings.
In practical terms, this consistency might reduce the need for human intervention when applying automated refactoring types.}

\rev{Furthermore, it is important to highlight that both models were able to present a high success rate for more unconventional refactorings, the ones that require design-level restructuring.
For example, DeepSeek could handle 22 unconventional refactorings, including \textit{pull up constructor body} and \textit{replace constructor with factory function}.
Such a result highlights the potential of using LLMs to perform refactorings that extend beyond syntactic transformations that typically require human expertise.}

\subsubsection{Real Scenarios}
\rev{Table~\ref{table: compilation-results-ref-finder-dataset} summarizes the performance of \gptmini\ and DeepSeek across different instruction strategies in the Real Scenario dataset across our 5 runs with their associated standard deviations.  Compilation rate serves as the first indicator, reflecting the proportion of refactorings that result in syntactically valid code. The two other metrics, \textit{New Failed Tests} and \textit{New Test Errors}, are only applicable for code that successfully compiles, as they measure semantic preservation.}

\rev{In the \textit{Zero-Shot} strategy, where only the name of the refactoring is provided without any accompanying guideline, both models are able to generate compilable code. \gptmini\ achieves a higher compilation rate than DeepSeek, with a large number of failed tests (0.807 on average) and test errors (0.333). In contrast, the compilable code of DeepSeek under zero-shot instructions did not introduce new test failures or errors, indicating stronger semantic preservation despite the slightly lower compilation rate. This suggests that when no detailed guidance is available, the models tend to rely on prior knowledge from pretraining, which may lead to inconsistent refactoring outcomes. Introducing examples in the \textit{Two-Shot} setting provides the models with additional context and improves reliability to some extent by reducing the test failure and test errors on \gptmini\ and improving the compilation ratio on DeepSeek. However, the improvements are not uniform, and \gptmini\ continues to show residual instability in compilation ratio, indicating that example-driven guidance alone may be insufficient for consistent correctness. The Step-by-Step and Rule-Based strategies, which provide explicit guidelines or refactoring rules, appear more promising and show more stability in results over both LLMs. These strategies are closer to human-oriented documentation, describing either the procedure or the transformation rules that must be applied to conduct a refactoring.  The compilation performance of both models was relatively close  (0.44 vs. 0.41 and 0.41 vs 0.38). While \gptmini\ again showed higher compilation rates than DeepSeek, this advantage was offset by the introduction of test failures and errors but the failure ratio in \textit{Step-by-Step} learning is lower than providing no instruction like Zero-Shot learning. For DeepSeek, the \textit{Rule-based} strategy reduces the compilation ratio. DeepSeek maintained perfect test preservation across both strategies, suggesting more reliable correctness once compilation succeeds. }

\rev{Finally, \textit{Objective Learning} differs from all other settings in that it does not explicitly mention the target refactoring type, but instead provides a high-level description of the concept of refactoring and relies more on the model's training. This instruction yields relatively high compilation rates for both models, with few test failures and errors in \gptmini\ and the highest compilation rate for DeepSeek. However, because the model is not pointed to a specific refactoring type, the resulting improvements may not consistently reflect true refactoring. Instead, they may stem from minimal changes—such as variable renaming—that preserve semantics without meaningfully improving code quality. Thus, while this strategy appears effective in generating compilable and test-preserving code, its impact on actual quality of code requires further investigation, which we address in RQ2. However, in the results collected for the \textit{Benchmark} dataset, human evaluation of success rate adopts a broader definition of correctness than only compilation or passing test suites. Human annotators compared the LLM-generated refactored code against the ground truth from Fowler’s catalog to ensure that the intended refactoring purpose was actually achieved. Consequently, within the \textit{Benchmark} dataset, \textit{Objective Learning} shows the lowest performance among the other strategies. }

\rev{The per-type analysis in Table~\ref{table: ref-finder-dataset-per-type} highlights notable differences across refactoring types. Some refactorings, such as \textit{Split Variable} and \textit{Extract Variable}, achieve consistently high compilation success for both models. DeepSeek compiles all cases of \textit{Split Variable} correctly (1.0), while \gptmini\ performs closely (0.967). Both models also achieve high rates on \textit{Extract Variable} (0.940 for DeepSeek and 0.910 for \gptmini), with no new failed tests or errors. These results suggest that variable-level refactorings, which involve localized transformations, are comparatively easier for LLMs to apply reliably, regardless of the provided instructions. Other refactorings, such as \textit{Change Function Declaration} and \textit{Slide Statements}, exhibit limited success across both models, with compilation rates below 0.4 for \gptmini\ and even lower for DeepSeek. Notably, \textit{Inline Variable} and \textit{Introduce Special Case} are the weakest points for both models, with neither generating compilable code in any instance. This suggests that certain refactoring types remain difficult for LLMs regardless of instruction strategy, possibly due to their dependence on broader contextual and design-level information. 
A contrast between the two models appears in \textit{Replace Nested Conditional with Guard Clauses}. GPT-4o m. achieves a compilation rate of 0.506, but this comes at the expense of semantic correctness. DeepSeek, in contrast, has a lower compilation rate for this refactoring (0.307), but the number of new test failures (0.033) and errors (0.0) is lower than \gptmini. 
Finally, \textit{Introduce Assertion} shows a limited success rate on \gptmini\ (0.200) while DeepSeek fails to compile any cases. This highlights that \gptmini, despite being more error-prone in refactoring code using different instructions, can sometimes outperform DeepSeek in comprehending less common refactoring tasks.}

\rev{Further analyzing the results per refactoring type, we observed that 
\textit{Replace Nested Condition with Guard Clauses} introduced a substantially higher 
number of new test failures compared to other types (e.g., 
3.388$\pm$8.861 for \gptmini). 
At the level of instruction strategies (Table~\ref{table: compilation-results-ref-finder-dataset}), these failures are averaged together with simpler refactorings, which explains why the strategy-level averages remain low despite the high variance in certain refactoring types. 
Therefore, results on test failures should be interpreted with caution, as they may be disproportionately influenced by a small number of semantically challenging refactoring types.}

\rev{Overall, the results on the Real Scenario dataset show that while both \gptmini\ and DeepSeek are capable of performing certain refactorings reliably—particularly localized transformations such as Split Variable and Extract Variable—their effectiveness varies across instruction strategies and refactoring types. \gptmini\ generally achieves higher compilation rates, but this advantage often comes with a higher ratio of semantic errors, as seen in complex transformations like Replace Nested Conditional with Guard Clauses. DeepSeek, on the other hand, exhibits lower compilation coverage but consistently stronger semantic preservation once code compiles, producing fewer test failures and errors across instruction styles.}

\begin{table}[t]
	\centering
        \caption{Compilation and Semantic Preservation Analysis across Instruction Types on Real Scenarios}
        \label{table: compilation-results-ref-finder-dataset}
	\resizebox{\textwidth}{!}{
		\begin{tabular}{|c|c|c|c|c|}
			\hline
			\textbf{LLM}                         & \textbf{Instruction Strategy}                     & \textbf{Avg. Compilation} & \textbf{Avg. New Failed Tests} & \textbf{Avg. New Test Errors} \\ \hline
			\multirow{5}{*}{\textbf{\gptmini}} 
			                                     & Zero-Shot Learning                        & 0.511$\pm$0.482          & 0.807$\pm$5.366            & 0.333$\pm$2.236                               \\ \cline{2-5} 
			                                     & Two-Shot Learning                              & 0.390$\pm$0.494              & 0.176$\pm$1.124        & 0.073$\pm$0.469                               \\ \cline{2-5} 
			                                     & Step-by-Step Learning                             & 0.443$\pm$0.493              & 0.171$\pm$1.111          & 0.071$\pm$0.463                            \\ \cline{2-5} 
			                                     & Rule-based Learning                             & 0.415$\pm$0.480              & 0.190$\pm$1.126                   & 0.073$\pm$0.469                  \\ \cline{2-5} 
			                                     & Objective Learning      & 0.443$\pm$0.475              & 0.004$\pm$0.029             & 0.009$\pm$0.059                               \\ \hline\hline
			\multirow{5}{*}{\textbf{DeepSeek}}     
			                                     & Zero-Shot Learning                          & 0.386$\pm$0.489                            & 0                                 & 0                                \\ \cline{2-5} 
			                                     & Two-Shot Learning                              & 0.410$\pm$0.494              & 0                                 & 0                                \\ \cline{2-5} 
			                                     & Step-by-Step Learning                              & 0.410$\pm$0.494             & 0                                 & 0                                \\ \cline{2-5} 
			                                     & Rule-based Learning                 & 0.380$\pm$0.483              & 0                                 & 0                                \\ \cline{2-5} 
			                                     & Objective Learning                      & 0.465$\pm$0.497                         & 0.060$\pm$0.244                              & 0                                \\ \hline
		\end{tabular}}
\end{table}

\begin{table}[]
	\centering
	\caption{Compilation Data and Semantic Analysis Per Refactoring Type on Real Scenarios
    }
	\label{table: ref-finder-dataset-per-type}
	\resizebox{1\textwidth}{!}{
		\begin{tabular}{|c|c|c|c|c|}
			\hline
			\textbf{LLM}                         & \textbf{Refactoring Type}                     & \textbf{Avg. Compilation} & \textbf{Avg. New Failed Tests} & \textbf{Avg. New Test Errors} \\ \hline
			\multirow{11}{*}{\textbf{\gptmini}} & Split Variable                                & 0.967$\pm$0.082              & 0                                 & 0                                \\ \cline{2-5} 
			                                     & Extract Variable                              & 0.910$\pm$0.279              & 0                                 & 0                                \\ \cline{2-5} 
			                                     & Extract Function                              & 0.611$\pm$0.473              & 0                                 & 0                                \\ \cline{2-5} 
			                                     & Replace Nested Cond. with Guard Clauses       & 0.506$\pm$0.480              & 
                                                 3.388$\pm$8.861                   & 1.412$\pm$3.692                  \\ \cline{2-5} 
			                                     & Replace Function with Command                 & 0.480$\pm$0.510              & 0                                 & 0                                \\ \cline{2-5} 
			                                     & Consolidate Cond. Expression                  & 0.416$\pm$0.493              & 0.008$\pm$0.040                   & 0.016$\pm$0.080                  \\ \cline{2-5} 
			                                     & Change Function Declaration                   & 0.352$\pm$0.452              & 0                                 & 0                                \\ \cline{2-5} 
			                                     & Introduce Assertion                           & 0.200$\pm$0.400              & 0                                 & 0                                \\ \cline{2-5} 
			                                     & Slide Statements                              & 0.183$\pm$0.371              & 0.026$\pm$0.113                   & 0                                \\ \cline{2-5} 
			                                     & Inline Variable                               & 0                            & 0                                 & 0                                \\ \cline{2-5} 
			                                     & Introduce Special Case                        & 0                            & 0                                 & 0                                \\ \hline\hline
			\multirow{11}{*}{\textbf{DeepSeek}}     & Split Variable                                & 1                            & 0                                 & 0                                \\ \cline{2-5} 
			                                     & Extract Variable                              & 0.940$\pm$0.226              & 0                                 & 0                                \\ \cline{2-5} 
			                                     & Extract Function                              & 0.676$\pm$0.475              & 0                                 & 0                                \\ \cline{2-5} 
			                                     & Replace Function with Command                 & 0.407$\pm$0.494              & 0                                 & 0                                \\ \cline{2-5} 
			                                     & Consolidate Cond. Expression                  & 0.400$\pm$0.500              & 0                                 & 0                                \\ \cline{2-5} 
			                                     & Replace Nested Cond. with Guard Clauses       & 0.307$\pm$0.453              & 0.033$\pm$0.129                     & 0                                \\ \cline{2-5} 
			                                     & Change Function Declaration                   & 0.224$\pm$0.410              & 0.020$\pm$0.100                   & 0                                \\ \cline{2-5} 
			                                     & Slide Statements                              & 0.155$\pm$0.360              & 0.045$\pm$0.251                   & 0                                \\ \cline{2-5} 
			                                     & Introduce Special Case                        & 0                            & 0                                 & 0                                \\ \cline{2-5} 
			                                     & Inline Variable                               & 0                            & 0                                 & 0                                \\ \cline{2-5} 
			                                     & Introduce Assertion                           & 0                            & 0                                 & 0                                \\ \hline
		\end{tabular}}
\end{table}

\subsection{\rev{RQ2: What is the impact of different instructions on the quality of code refactored by LLM?}}
\rev{While RQ1 aimed to evaluate the LLMs' capability to successfully apply a wide range of refactoring types considering different types of instructions, RQ2 focuses on investigating the quality of refactored code by the LLM.}
\rev{For that, we assess the quality of refactorings produced by LLMs using several metrics (e.g., LOC, CC, CodeBLEU). These metrics are computed relative to the ground-truth refactored code, with higher CodeBLEU scores indicating higher similarity and reductions in LOC, CC, and FOUT reflecting improvements in conciseness and complexity. We first present results on the benchmark dataset, followed by real scenarios.
}

\subsubsection{Benchmark Scenarios}

\begin{table}[t]
	\centering
	\caption{Quality Metrics per Instruction Strategy (Benchmark). Before and after values correspond to the benchmark code; when comparing the results, we rely only on the after-refactoring scores (ground truth).} 
	\label{table: fowler-dataset-quality}
	\resizebox{\textwidth}{!}{
		\begin{tabular}{|cc|c|c|c|c|}
			\hline
			\multicolumn{2}{|c|}{\textbf{}}                                                    & \textbf{CodeBLEU} & \textbf{CC} & \textbf{LOC} & \textbf{FOUT} \\ \hline
			\multicolumn{2}{|c|}{Before Refactoring}                    &                   & 0.443$\pm$1.103       & 3.574$\pm$3.243                  & 0.557$\pm$1.025                 \\ \hline
			\multicolumn{2}{|c|}{After Refactoring (Ground Truth)}                     &                   & 0.279$\pm$0.773       & 3.590$\pm$3.247                  & 0.836$\pm$1.280                 \\ \hline \hline 
			
			\multicolumn{1}{|c|}{\multirow{4}{*}{\gptmini}} & Zero-Shot Learning       & 0.303$\pm$0.132 & 0.348$\pm$0.838 & 11.472$\pm$8.036  & 1.518$\pm$1.987 \\ \cline{2-6} 
			\multicolumn{1}{|c|}{}                              & Step-by-Step Learning    & 0.359$\pm$0.151 & 0.479$\pm$1.108 & 14.639$\pm$11.098 & 1.898$\pm$2.109 \\ \cline{2-6} 
			\multicolumn{1}{|c|}{}                              & Rule-based Learning      & 0.359$\pm$0.144 & 0.352$\pm$0.867 & 11.100$\pm$7.824  & 1.243$\pm$1.742 \\ \cline{2-6} 
			\multicolumn{1}{|c|}{}                              & Objective Learning       & 0.278$\pm$0.130 & 0.459$\pm$1.066 & 11.338$\pm$7.262  & 1.692$\pm$2.148 \\ \hline \hline
            
			\multicolumn{1}{|c|}{\multirow{4}{*}{DeepSeek}}     & Zero-Shot Learning       & 0.359$\pm$0.179 & 0.311$\pm$0.827 & 10.941$\pm$9.364  & 1.607$\pm$2.173 \\ \cline{2-6} 
			\multicolumn{1}{|c|}{}                              & Step-by-Step Learning    & 0.388$\pm$0.191 & 0.489$\pm$1.149 & 13.852$\pm$11.655 & 2.000$\pm$2.375 \\ \cline{2-6} 
			\multicolumn{1}{|c|}{}                              & Rule-based Learning      & 0.407$\pm$0.185 & 0.513$\pm$1.026 & 11.409$\pm$9.382  & 1.400$\pm$1.854 \\ \cline{2-6} 
			\multicolumn{1}{|c|}{}                              & Objective Learning       & 0.296$\pm$0.149 & 0.311$\pm$0.748 & 9.416$\pm$6.763   & 1.567$\pm$2.071 \\ \hline
		\end{tabular}}
\end{table}

Table \ref{table: fowler-dataset-quality} presents the average metrics collected across five runs from Fowler's benchmark, aggregated by instruction type, along with the associated standard deviations. \rev{It is worth noting that some code snippets in the \textit{Benchmark} dataset are incomplete, as further discussed in Section~\ref{sec:threats}. In such cases, when LLMs attempted to refactor the code, they also tried to complete the missing parts, which resulted in an increase in LOC compared to the ground truth that can consequently impacts the CC and FOUT. Thus, in Table \ref{table: fowler-dataset-quality}, while comparing the CC, LOC, and FOUT results across different instruction types provides insight into their relative usefulness for guiding refactoring tasks, comparing these values directly with the ground truth may be inconclusive.}

\rev{Regarding CodeBLEU, we observe that all instruction strategies result in low similarity between the generated code and the ground truth (0.296-0.407). Among them, Rule-based Learning (0.359 for GPTmini, 0.407 for DeepSeek), followed by Step-by-Step Learning (0.359 for GPTmini, 0.388 for DeepSeek), achieves the highest scores, indicating that these strategies generate refactorings most similar to human-applied transformations. In contrast, Objective Learning yields much lower similarity (0.278 for GPTmini, 0.296 for DeepSeek). As discussed earlier, this can be explained by the fact that Objective Learning does not force the model to apply a specific transformation required by the ground truth; instead, it allows the model to refactor code with the broader goal of improving quality while preserving behavior. Interestingly, when applying Objective Learning, DeepSeek achieves the lowest CC (0.311) and LOC (9.416), while GPTmini records the third-lowest CC (0.459) and the second-lowest LOC (11.338), following Rule-based Learning (11.1 LOC, 0.352 CC).}

\rev{For \gptmini, step-by-step instructions result in the largest LOC (14.639) and FOUT (1.898), while rule-based instructions keep both values lower (11.1 LOC and 1.243 FOUT). Objective Learning generates values close to zero-shot and rule-based strategies but slightly higher FOUT (1.692). In terms of CC, all strategies remain close, ranging from 0.348 to 0.479, with step-by-step producing the highest CC (0.479). DeepSeek shows a similar spread: step-by-step instructions again generate the largest LOC (13.852) and FOUT (2.0), whereas objective learning yields the smallest LOC (9.416). Rule-based instructions have the second lowest LOC (11.409) with relatively low FOUT (1.400). CC values for DeepSeek also vary within a narrow range (0.311–0.513), with step-by-step and rule-based slightly higher than zero-shot and objective learning.}

\rev{Since the benchmark dataset covers 61 refactoring types, we did not include a type-by-type analysis here due to space limitations. Instead, we present the averaged scores across all five runs for each refactoring type in our replication package \citep{appendix} and we only discuss the main insights here. The refactoring type analysis highlights that both \gptmini\ and DeepSeek achieve better results (e.g., lower CC) on relatively simple, localized transformations such as \textit{Rename Variable}, \textit{Split Variable}, and \textit{Replace Nested Condition with Guard Clauses}. These refactorings require little context beyond the method itself. DeepSeek further extends this success to more types, including \textit{Pull Up Constructor Body}, \textit{Extract Superclass}, and \textit{Extract Variable}. In contrast, both models perform poorly (e.g., higher CC or higher LOC) on more complex transformations such as \textit{Decompose Conditional}, \textit{Combine Functions Into Transform}, and \textit{Encapsulate Variable}. These cases often led the models to generate additional method implementations. Considering \textit{Replace Nested Condition with Guard Clauses} illustrates that although both models produced similar CC and FOUT values, DeepSeek generated significantly lower LOC. }

\subsubsection{Real Scenarios}

\begin{table}[t]
	\centering
        \caption{Quality Metrics per Prompt Strategy (Real Scenario). Before and after values correspond to the original code and the developer-applied refactorings; when comparing the results, we rely only on the after-refactoring scores (ground truth).}
        \label{table: ref-finder-dataset}
	\resizebox{\textwidth}{!}{
		\begin{tabular}{|cc|c|c|c|c|}
			\hline
			\multicolumn{2}{|c|}{\textbf{}}                                                    & \textbf{CodeBLEU} & \textbf{CC} & \textbf{LOC} & \textbf{FOUT} \\ \hline
			\multicolumn{2}{|c|}{Before Refactoring}                    &                 & 6.081$\pm$5.675       & 23.735$\pm$17.412                & 10.459$\pm$9.279                 \\ \hline
			\multicolumn{2}{|c|}{After Refactoring (Ground Truth)}                     &                 & 5.820$\pm$5.654       & 23.741$\pm$16.586                & 11.346$\pm$10.028                 \\ \hline \hline
			
			\multicolumn{1}{|c|}{\multirow{5}{*}{\gptmini}} 
                & Zero-Shot Learning       & 0.558$\pm$0.078 & 5.986$\pm$5.446 & 28.050$\pm$18.480 & 11.455$\pm$10.039 \\ \cline{2-6}
			\multicolumn{1}{|c|}{} & Two-Shot Learning       & 0.597$\pm$0.094 & 6.120$\pm$5.672 & 26.981$\pm$17.112 & 10.753$\pm$9.170 \\ \cline{2-6}
			\multicolumn{1}{|c|}{} & Step-by-Step Learning   & 0.589$\pm$0.076 & 6.008$\pm$5.418 & 27.445$\pm$18.715 & 11.124$\pm$9.831 \\ \cline{2-6} 
			\multicolumn{1}{|c|}{} & Rule-based Learning     & 0.597$\pm$0.087 & 6.032$\pm$5.624 & 25.163$\pm$15.459 & 10.599$\pm$8.981 \\ \cline{2-6} 
			\multicolumn{1}{|c|}{} & Objective Learning      & 0.521$\pm$0.052 & 5.825$\pm$5.168 & 31.359$\pm$24.972 & 12.193$\pm$10.749 \\ \hline \hline
            
			\multicolumn{1}{|c|}{\multirow{5}{*}{DeepSeek}}     
                & Zero-Shot Learning       & 0.541$\pm$0.060 & 5.880$\pm$5.495 & 31.803$\pm$18.794 & 11.370$\pm$9.092 \\ \cline{2-6}
			\multicolumn{1}{|c|}{} & Two-Shot Learning       & 0.534$\pm$0.062 & 6.115$\pm$5.609 & 31.909$\pm$19.379 & 10.854$\pm$8.657 \\ \cline{2-6}
			\multicolumn{1}{|c|}{} & Step-by-Step Learning   & 0.530$\pm$0.053 & 6.253$\pm$5.576 & 32.018$\pm$18.426 & 11.217$\pm$9.480 \\ \cline{2-6} 
			\multicolumn{1}{|c|}{} & Rule-based Learning     & 0.534$\pm$0.059 & 6.009$\pm$5.234 & 32.144$\pm$20.061 & 11.101$\pm$9.325 \\ \cline{2-6} 
			\multicolumn{1}{|c|}{} & Objective Learning      & 0.539$\pm$0.054 & 6.052$\pm$5.615 & 27.957$\pm$21.200 & 11.544$\pm$10.417 \\ \hline
		\end{tabular}}
\end{table}

\rev{Table \ref{table: ref-finder-dataset} presents the quality metrics obtained from different instruction strategies and provides a comparison against the ground truth baseline before and after refactoring. 
In fact, most strategies yield CodeBLEU values below 0.6, which can support the interpretation that both \gptmini\ and DeepSeek are generating novel solutions rather than only reproducing answers from their training data that can mitigate concerns about memorization in LLMs. \gptmini\ shows greater variability across instructions, ranging from 0.521 under \textit{Objective Learning} to 0.597 under \textit{Two-Shot} and \textit{Rule-based Learning}. By contrast, code refactored by DeepSeek produces more consistent scores (0.530–0.541) across instructions, though generally are less similar to ground truth than \gptmini.}

\rev{In general, the CodeBLEU scores observed for both models across all instruction strategies show a moderate similarity with human-written code (ground truth after refactoring). 
This suggests that varying the instruction strategy had little to no impact on how closely the LLMs' outputs resembled human code.}

\rev{When examining cyclomatic complexity (CC), the ground truth shows a reduction after refactoring (from 6.08 to 5.82), which can be considered as an expected outcome of refactoring. Both models remain close to this baseline in CC, with \gptmini\ \textit{Objective Learning} yielding the lowest CC (5.82), slightly outperforming the ground truth after refactoring. By providing only the overall objective of refactoring—without prescribing any particular refactoring type—both LLMs reduced CC in the refactored code as an indicator of improving code quality. Other strategies, particularly Two-Shot for \gptmini\ and Step-by-Step for DeepSeek, slightly increase CC, suggesting that in some cases refactoring instructions may lead to more complex control flow rather than simplification, which can also be attributed to greater LOC.}

\rev{The average of LOC in ground truth remains stable before and after refactoring (~23.7), but both models consistently generate longer code compared to ground truth. \gptmini\ produces LOC  between 25.16 and 31.36, with \textit{Rule-based Learning} being the closest to baseline and \textit{Objective Learning} resulting in a considerable increase. DeepSeek, with LOC ranging from 27.96 to 32.14 across strategies, indicating that its refactorings often introduce additional code structure or scaffolding that can also impact the CC of the generated code.}

\rev{The average of FOUT, which reflects the number of method calls, increases on ground truth after refactoring, which can be attributed to a more modular code such as generating the new method to apply a specific refactoring type. This metric also increases relative to ground truth across different instruction strategies with \textit{Step-by-Step} and \textit{Zero-shot} Learning closer to the ground truth. For example, \gptmini, Rule-based Learning produces the lowest FOUT (10.60), close to the baseline before refactoring, whereas \textit{Objective Learning} leads to the highest (12.19). These results suggest that certain instruction strategies may encourage LLMs to generate code that is more modular or decomposed into smaller calls, whereas others lead to designs that rely on additional external methods.}

\begin{table}[htbp]
    \caption{Quality Metrics per Refactoring Type (Real Scenarios). Before and after values correspond to the original code and the developer-applied refactorings (ground truth).}
	\label{table: refactoring-type-metrics}
	\resizebox{1.0\textwidth}{!}{
		\begin{tabular}{|c|c|c|c|c|c|}
			\hline
			\textbf{LLM}                & \textbf{Refactoring Type}                       & \textbf{CodeBleu} & \textbf{CC} & \textbf{LOC} & \textbf{FOUT} \\ \hline

			                              & Change Function Declaration                     &    & 4.333$\pm$6.144   & 18.556$\pm$25.559             & 6.667$\pm$9.734             \\ \cline{2-6} 
			                              & Consolidate Conditional Expression              &    & 7.600$\pm$6.580   & 27.600$\pm$20.120             & 12.000$\pm$11.225            \\ \cline{2-6} 
			                              & Extract Function                                &    & 3.857$\pm$5.113   & 17.429$\pm$12.109             & 9.000$\pm$5.323            \\ \cline{2-6} 
		                                  & Extract Variable                                &    & 1.000$\pm$2.000       & 9.000$\pm$6.164             & 5.750$\pm$6.449             \\ \cline{2-6}
			                              & Inline Variable                                 &    & 1.000$\pm$0      & 8.000$\pm$0           & 3.000$\pm$0                \\ \cline{2-6} 
			                              & Introduce Assertion                             &    & 0.333$\pm$0.577   & 10.000$\pm$5.568             & 6.667$\pm$4.041
             \\ \cline{2-6} 
			                              & Introduce Special Case                          &    & 18.000$\pm$0  & 56.000$\pm$0            & 23.000$\pm$0            \\ \cline{2-6} 
			                              & Replace Function With Command                   &    & 6.167$\pm$6.824 & 15.000$\pm$14.057 & 2.167$\pm$1.329             \\ \cline{2-6} 
			                              & Replace Nested Conditional With Guard Clauses   &    & 13.000$\pm$5.033  & 45.500$\pm$12.124             & 15.000$\pm$8.963            \\ \cline{2-6} 
			                              & Slide Statements                                &    & 10.100$\pm$5.801  & 45.500$\pm$23.287             & 30.800$\pm$26.587            \\ \cline{2-6} 
		   \multirow{-11}{*}{\shortstack{Before Refactoring}}   & Split Variable                                &    & 1.500$\pm$2.121       & 8.500$\pm$2.121             & 1.000$\pm$1.414             \\ \hline \hline 

			                              & Change Function Declaration                     &    & 4.667$\pm$7.467   & 22.222$\pm$29.995             & 7.444$\pm$12.095             \\ \cline{2-6} 
			                              & Consolidate Conditional Expression              &    & 8.000$\pm$6.819   & 29.000$\pm$19.274             & 12.600$\pm$10.831            \\ \cline{2-6} 
			                              & Extract Function                                &    & 4.000$\pm$4.546   & 23.429$\pm$12.934            & 10.714$\pm$5.282            \\ \cline{2-6} 
		                                  & Extract Variable                                &    & 1.750$\pm$1.708       & 14.000$\pm$6.481             & 7.250$\pm$5.795             \\ \cline{2-6}
			                              & Inline Variable                                 &    & 1.000$\pm$0      & 7.000$\pm$0           & 2.000$\pm$0                \\ \cline{2-6} 
			                              & Introduce Assertion                             &    & 0.333$\pm$0.577   & 11.000$\pm$5.568             & 7.667$\pm$4.509             \\ \cline{2-6} 
			                              & Introduce Special Case                          &    & 17.000$\pm$0  & 48.000$\pm$0            & 25.000$\pm$0            \\ \cline{2-6} 
			                              & Replace Function With Command                   &    & 1.667$\pm$4.082 & 7.000$\pm$8.556 & 1.833$\pm$1.329             \\ \cline{2-6} 
			                              & Replace Nested Conditional With Guard Clauses   &    & 13.000$\pm$5.657  & 43.000$\pm$7.071             & 16.500$\pm$12.021            \\ \cline{2-6} 
			                              & Slide Statements                                &    & 11.100$\pm$6.574  & 49.500$\pm$27.342             & 32.800$\pm$27.720            \\ \cline{2-6} 
		   \multirow{-11}{*}{\shortstack{After Refactoring\\(Ground Truth))}}  & Split Variable                                &    & 1.500$\pm$2.121       & 7.000$\pm$2.828             & 1.000$\pm$1.414             \\ \hline \hline 

			                              & Introduce Assertion                             & 0.667$\pm$0.065   & 0.440$\pm$0.121   & 13.493$\pm$1.175             & 7.227$\pm$0.358             \\ \cline{2-6} 
			                              & Introduce Special Case                          & 0.658$\pm$0.043   & 17.320$\pm$0.743  & 67.480$\pm$12.032            & 26.040$\pm$3.160            \\ \cline{2-6} 
			                              & Split Variable                                  & 0.640$\pm$0.048   & 1.500$\pm$0       & 12.280$\pm$4.160             & 1.360$\pm$0.699             \\ \cline{2-6} 
			                              & Extract Function                                & 0.596$\pm$0.044   & 3.937$\pm$0.175   & 23.634$\pm$1.564             & 10.171$\pm$0.326            \\ \cline{2-6} 
			                              & Replace Nested Conditional With Guard Clauses   & 0.595$\pm$0.042   & 12.480$\pm$0.444  & 46.620$\pm$5.330             & 15.580$\pm$1.411            \\ \cline{2-6} 
			                              & Slide Statements                                & 0.592$\pm$0.106   & 10.144$\pm$0.062  & 48.888$\pm$6.687             & 31.628$\pm$1.508            \\ \cline{2-6} 
			                              & Consolidate Conditional Expression              & 0.557$\pm$0.050   & 7.648$\pm$0.212   & 27.528$\pm$2.854             & 12.040$\pm$0.616            \\ \cline{2-6} 
			                              & Replace Function With Command                   & 0.536$\pm$0.022   & 6.187$\pm$0.198   & 24.267$\pm$6.853             & 2.860$\pm$0.155             \\ \cline{2-6} 
			                              & Change Function Declaration                     & 0.518$\pm$0.019   & 4.280$\pm$0.073   & 20.475$\pm$1.941             & 6.956$\pm$0.447             \\ \cline{2-6}
                                            & Inline Variable                                 & 0.496$\pm$0.023   & 1.000$\pm$0       & 9.440$\pm$0.984              & 3.080$\pm$0.179             \\ \cline{2-6}
		  \multirow{-11}{*}{\gptmini}   & Extract Variable                                & 0.442$\pm$0.021   & 1.000$\pm$0       & 11.690$\pm$1.118             & 6.530$\pm$0.347             \\ \hline \hline 

		                                  & Split Variable                                  & 0.620$\pm$0.021   & 1.500$\pm$0       & 12.080$\pm$2.027             & 1.240$\pm$0.428             \\ \cline{2-6}
			                              & Introduce Special Case                          & 0.582$\pm$0.017   & 16.800$\pm$1.049  & 69.600$\pm$4.626             & 24.600$\pm$2.025            \\ \cline{2-6} 
			                              & Replace Nested Conditional With Guard Clauses   & 0.580$\pm$0.019   & 13.420$\pm$0.559  & 52.220$\pm$0.998             & 16.100$\pm$0.686            \\ \cline{2-6} 
			                              & Introduce Assertion                             & 0.579$\pm$0.031   & 0.480$\pm$0.159   & 16.693$\pm$2.547             & 7.453$\pm$0.568             \\ \cline{2-6} 
			                              & Slide Statements                                & 0.549$\pm$0.030   & 10.064$\pm$0.288  & 49.708$\pm$2.479             & 30.828$\pm$1.819            \\ \cline{2-6} 
			                              & Extract Function                                & 0.547$\pm$0.021   & 4.086$\pm$0.169   & 28.423$\pm$4.748             & 10.531$\pm$0.866            \\ \cline{2-6} 
			                              & Consolidate Conditional Expression              & 0.516$\pm$0.022   & 8.296$\pm$0.582   & 33.400$\pm$2.860             & 13.032$\pm$0.648            \\ \cline{2-6} 
			                              & Replace Function With Command                   & 0.498$\pm$0.028   & 5.880$\pm$0.939   & 33.913$\pm$11.643            & 3.167$\pm$1.056             \\ \cline{2-6} 
			                              & Inline Variable                                 & 0.497$\pm$0.020   & 1.000$\pm$0       & 10.080$\pm$0.782           & 3.000$\pm$0                 \\ \cline{2-6} 
			                              & Change Function Declaration                     & 0.479$\pm$0.029   & 4.151$\pm$0.166   & 24.040$\pm$2.935             & 6.849$\pm$0.338             \\ \cline{2-6} 
		   \multirow{-11}{*}{DeepSeek}   & Extract Variable                                & 0.443$\pm$0.011   & 1.000$\pm$0       & 12.670$\pm$1.281             & 6.590$\pm$0.216             \\ \hline 

		\end{tabular}}
\end{table}

Moving forward, Table \ref{table: refactoring-type-metrics} presents the average quality metrics per refactoring type. 
\rev{The ground truth corresponds to the human-validated version of the code after applying the refactoring. Each LLM’s results are compared against this ground truth to evaluate whether the refactored code improves quality metrics relative to the original version.}\footnote{We report both before and after values for the ground truth so that the improvement produced by human-applied refactorings can be directly compared with LLM-applied refactorings.}
\rev{Starting with CodeBLEU, we observe that LLM-generated code snippets differ from ground truth code, ranging from moderate to low similarity.}
Analyzing the CodeBLEU score of the \textit{Extract Variable} refactoring, one of the most successful refactoring performed by both LLMs, we observe that while LLMs do not introduce many cases with compilation issues, or tests with failures or errors (see Table~\ref{table: ref-finder-dataset-per-type}), LLMs typically apply this refactoring type differently from human developers. 
Specifically, this refactoring reported the lowest CodeBLEU scores on both models (0.442 and 0.443 for \gptmini and DeepSeek, respectively).  
On the contrary, \textit{Introduce Assertion} and \textit{Introduce Special Case}, which produced two of the lowest average compilation rates, averaged some of the highest CodeBLEU scores across both models (0.667 and 0.658, and 0.620 and 0.528, for \gptmini and DeepSeek, respectively). 

\rev{Regarding the remaining metrics, we observed that LLM-generated snippets shared similar outcomes. 
For example, for CC and FOUT, although some differences are present, their overall scores are comparable, highlighting consistent behavior across these complexity metrics.
Compared to the ground truth, we observed that in most cases, the solutions generated by the LLMs exhibited lower Cyclomatic Complexity (CC), not statistically significant.\footnote{Mann-Whitney, p-value = \ 0.0638 } 
However, this trend did not hold across all refactoring types. 
For instance, in the case of the \textit{Replace Function with Command} refactoring, the CC scores were more than three times higher than those observed in the ground truth implementations.
For LOC, we observed that DeepSeek was more verbose when compared to \gptmini.
Compared with the ground truth, we observed that for most cases, DeepSeek required more lines of code to address the required refactoring, leading to high scores, like \textit{Introduce Special Case} (69.6), even not statistically significant.\footnote{Mann-Whitney, p-value = \ 0.481}
Such increased verbosity could indicate more detailed implementations or additional scaffolding code, which might improve readability or functionality but could also lead to higher maintenance overhead. 
}

\section{Discussion}
\label{sec:discussion}

In this section, we further discuss our results, focusing on their implications and future improvements.

\subsection{Discovering LLMs' Capacity for Refactorings}
Overall, our findings suggest that LLMs demonstrate capabilities in performing code refactorings.
Notably, the evaluated models successfully handled a broad range of refactoring types, including some that are not even supported by existing assistant tools.
As such, our results reinforce previous studies that place LLMs as a promising new assistant tool supporting practitioners in applying refactorings during their daily tasks.

Regarding the LLMs under analysis, \gptmini\ and DeepSeek exhibit similar strengths and challenges in certain directions.
For instance, both models struggle with the same refactoring types in our sample of Benchmark's scenarios. 
When evaluating our sample of real scenarios, some refactorings led to more problematic solutions, such as non-compilable code or failing/errored tests.
Given that preserving behavior is a core principle of refactoring, these challenges highlight the need for further improvements in LLM-generated refactorings, particularly in ensuring both syntactic and semantic correctness. \rev{However, the two models show different outcomes depending on the instruction strategy. For instance, under Zero-Shot learning—where only the name of the refactoring type is provided without additional guidance—DeepSeek performs better than \gptmini~. In contrast, \gptmini\ struggles to successfully perform diverse range of refactorings without additional explanations clarifying the requirements of each transformation. This difference could be attributed to variations in their training data, with one model having stronger prior knowledge of refactoring types. }

\rev{We also examined an instruction strategy called Objective Learning, which provides only the overall objective and purpose of refactoring without specifying a particular refactoring type. Our results show that, while LLMs often failed to correctly apply the intended refactoring type under this strategy, the final outputs nonetheless exhibited better performance on code quality metrics. This suggests that, without being constrained by detailed guidelines or forced into a specific refactoring type, LLMs can still improve code quality by focusing on the broader objective of refactoring.}

\rev{This finding highlights the value of aligning with Fowler’s definition of refactoring—\textit{“the process of changing source code to improve readability, maintainability, and quality without altering its external behavior".} By granting LLMs the freedom to pursue this objective rather than prescribing a specific transformation, the resulting code can achieve higher quality (e.g., reduced cyclomatic complexity) compared to cases where the model is directed toward a particular refactoring type. Thus, the choice of instruction strategy should depend on the task’s purpose: if the goal is to enforce a specific transformation and refactoring type, explicit instructions are necessary and the instruction strategy can be selected considering the core LLM; otherwise, providing the general objective may allow the model to apply its own strategy and still achieve the desired quality improvements.}

\rev{Comparing across instruction strategies, our results indicate that Rule-based learning yielded better performance across different refactoring types, particularly with \gptmini~. In contrast, for DeepSeek, instructions with minimal guidance—such as Zero-Shot—proved equally effective, and in some cases even outperformed other strategies for certain refactoring types which as discussed can be attributed to its training data. The key distinction lies in the fact that Rule-based instructions are derived from the rules used by automated refactoring tools to detect refactoring types, whereas Step-by-Step instructions are written as descriptive sequences of actions that are more easily understood by humans. Our results highlight equal or better performance in applying diverse range of refactorings while considering Rule-based strategy compared to pre-defined Step-by-Step learning. }

\subsection{Compilation Issues faced by LLMs}

\begin{table}[ht]
\resizebox{\textwidth}{!}{
\begin{tabular}{|c|c|c|c|c|c|c|c|}
\hline
\textbf{LLM} & \textbf{Project} & \textbf{Error - Phase} & \textbf{Run 1} & \textbf{Run 2} & \textbf{Run 3} & \textbf{Run 4} & \textbf{Run 5} \\ \hline
\multirow{9}{*}{\gptmini} 
 & \multirow{5}{*}{Antrl4}  & Total of attempts      & 179 & 186 & 186 & 179 & 179 \\ \cline{3-8}
 &                          & Lexical               & 48  & 47  & 47  & 48  & 48  \\ \cline{3-8}
 &                          & Parsing               & 8   & 0   & 0   & 3   & 3   \\ \cline{3-8}
 &                          & Compilation error     & 49  & 51  & 51  & 56  & 56  \\ \cline{3-8}
 &                          & Total Errors          & \shortstack{105 \\ (58.65\%)} & \shortstack{98 \\ (52.68\%)} & \shortstack{98 \\ (52.68\%)} & \shortstack{107 \\ (59.77\%)} & \shortstack{107 \\ (59.77\%)} \\ \cline{2-8}
 & \multirow{4}{*}{JUnit}   & Total of attempts     & 30  & 30  & 30  & 30  & 30  \\ \cline{3-8}
 &                          & Parsing               & 4   & 4   & 4   & 4   & 4   \\ \cline{3-8}
 &                          & Compilation error     & 2   & 2   & 2   & 2   & 2   \\ \cline{3-8}
 &                          & Total Errors          & \shortstack{6 \\ (20\%)} & \shortstack{6 \\ (20\%)} & \shortstack{6 \\ (20\%)} & \shortstack{6 \\ (20\%)} & \shortstack{6 \\ (20\%)} \\ \hline
\multirow{9}{*}{DeepSeek} 
 & \multirow{5}{*}{Antrl4}  & Total of attempts      & 179 & 179 & 179 & 179 & 179 \\ \cline{3-8}
 &                          & Lexical               & 48  & 48  & 48  & 48  & 48  \\ \cline{3-8}
 &                          & Parsing               & 8   & 5   & 5   & 5   & 5   \\ \cline{3-8}
 &                          & Compilation error     & 49  & 53  & 53  & 53  & 53  \\ \cline{3-8}
 &                          & Total Errors          & \shortstack{105 \\ (58.65\%)} & \shortstack{106 \\ (59.21\%)} & \shortstack{106 \\ (59.21\%)} & \shortstack{106 \\ (59.21\%)} & \shortstack{106 \\ (59.21\%)} \\ \cline{2-8}
 & \multirow{4}{*}{JUnit}   & Total of attempts     & 30  & 30  & 30  & 30  & 30  \\ \cline{3-8}
 &                          & Parsing               & 1   & 1   & 1   & 1   & 1   \\ \cline{3-8}
 &                          & Compilation error     & 4   & 4   & 4   & 4   & 4   \\ \cline{3-8}
 &                          & Total Errors          & \shortstack{5 \\ (16.66\%)} & \shortstack{5 \\ (16.66\%)} & \shortstack{5 \\ (16.66\%)} & \shortstack{5 \\ (16.66\%)} & \shortstack{5 \\ (16.66\%)} \\ \hline
\end{tabular}
}
\caption{Distribution of failed attempts to generate refactorings by LLMs}
\label{tab:parsing-errors-combined}
\end{table}

\rev{When using LLMs to perform the refactorings, we observe that some attempts resulted in failures, especially for our sample of real scenarios.
Aiming to gain a better understanding of these errors and further improve our approach, we provide a detailed analysis here.
Table \ref{tab:parsing-errors-combined} presents the distribution of errors across the different runs performed for \gptmini\ and DeepSeek, respectively.
We can observe three types of errors that occur at different stages of the compilation process (lexical, parsing, and compilation), all of which are detected by our scripts during the automatic validation step.
Here, we initially focus on the errors caught by our parser (lexical and parsing, see Section \ref{sec:automatic-validation}), later discussing the ones caught during the attempt to compile the code.}

\rev{First, we have the \textit{lexical errors}, which occur when a sequence of invalid tokens is encountered in a given programming language, in our case, Java.
For both LLMs, we observe a consistent number of such failures, even across different runs, with the proportion of failed attempts ranging from 25\% to 31\% for \gptmini\ and DeepSeek, respectively. 
For example, consider the code snippet presented in Listing \ref{error-lexical}, generated by DeepSeek for the Antlr4 project.
The error occurs because the string literal ends with a lone backslash (\textbackslash). 
However, in Java, this denotes the start of an escape sequence, but here it is immediately followed by a newline instead of a valid escape character. 
As a result, the lexical analyzer cannot tokenize the string correctly, leading to a lexical error.
Further analysis of code snippets generated by \gptmini\ revealed that LLMs failed in the same cases.
} 

\begin{lstlisting}[style=javastyle,caption={Excerpt with faulty string literal.}, label=error-lexical, floatplacement=t]
if (config.state instanceof RuleStopState) {
    if (debug) {
        if (recog != null) {
            System.out.format("closure at %s rule stop %s\
", recog.getRuleNames()[config.state.ruleIndex], config);
        } else {
            System.out.format("closure at rule stop %s\
", config); {...}
\end{lstlisting}

\begin{lstlisting}[style=javastyle,caption={Faulty generated method signature.},label=parsing-error,floatplacement=htbp]
public constructor TestCase(String name) or TestCase()
\end{lstlisting}

\rev{Second, we have \textit{parsing errors}, which occur when the sequence of tokens does not conform to the syntactic rules of the programming language.
This time, the LLMs report slightly different results when comparing the target projects. 
Although they exhibit similar frequencies of failed attempts across runs, \gptmini\ shows a higher number of failures on the JUnit project. 
For example, consider the code snippet reported in Listing \ref{parsing-error}, generated by \gptmini\ for the JUnit project.
The error arises because the generated method header contains the token \texttt{or}, which is not valid Java syntax in this context. 
After a parameter list, the Java grammar only permits tokens such as \texttt{;}, \texttt{@}, \texttt{[}, \texttt{throws}, or an opening brace \texttt{\{} for the method body. 
Since \texttt{or} does not match any expected token, the parser reports a syntax error.
}

\begin{table}[htbp]
\resizebox{\textwidth}{!}{
\begin{tabular}{|c|l|l|r|r|r|r|r|}
\hline
\textbf{LLM} & \textbf{Project} & \textbf{Error message} & \textbf{Run 1} & \textbf{Run 2} & \textbf{Run 3} & \textbf{Run 4} & \textbf{Run 5} \\ \hline

\multirow{18}{*}{\gptmini}
  & \multirow{16}{*}{Antlr4} & cannot find symbol: variable & 48 & 61 & 61 & 76 & 76 \\ \cline{3-8}
  &                          & is already defined in class & 27 & 26 & 26 & 29 & 29 \\ \cline{3-8}
  &                          & cannot find symbol: method & 29 & 13 & 13 & 31 & 31 \\ \cline{3-8}
  &                          & incompatible types & 11 & 13 & 13 & 11 & 11 \\ \cline{3-8}
  &                          & \begin{tabular}[c]{@{}l@{}}cannot be referenced from a\\ static context\end{tabular} & 5 & 10 & 10 & 8 & 8 \\ \cline{3-8}
  &                          & illegal parenthesized expression & 5 & 5 & 5 & 8 & 8 \\ \cline{3-8}
  &                          & cannot assign a value to final variable & 4 & 0 & 0 & 6 & 6 \\ \cline{3-8}
  &                          & cannot be accessed from outside package & 2 & 2 & 2 & 2 & 2 \\ \cline{3-8}
  &                          & Illegal static declaration in inner class & 1 & 3 & 3 & 1 & 1 \\ \cline{3-8}
  &                          & \begin{tabular}[c]{@{}l@{}}method does not override or implement\\ a method from a supertype\end{tabular} & 0 & 0 & 0 & 0 & 0 \\ \cline{3-8}
  &                          & diamond operator is not supported in -source 6 & 1 & 2 & 2 & 1 & 1 \\ \cline{3-8}
  &                          & cannot be applied to given types & 0 & 3 & 3 & 0 & 0 \\ \cline{3-8}
  &                          & missing return statement & 0 & 1 & 1 & 0 & 0 \\ \cline{3-8}
  &                          & modifier static not allowed here & 0 & 0 & 0 & 0 & 0 \\ \cline{3-8}
  &                          & \textbf{Total} & \textbf{134} & \textbf{141} & \textbf{141} & \textbf{174} & \textbf{174} \\ \cline{2-8}
  & \multirow{2}{*}{JUnit}   & cannot find symbol: method  & 2 & 2 & 2 & 2 & 2 \\ \cline{3-8}
  &                          & \textbf{Total} & \textbf{2} & \textbf{2} & \textbf{2} & \textbf{2} & \textbf{2} \\ \hline\hline

\multirow{18}{*}{DeepSeek}
  & \multirow{16}{*}{Antlr4} & cannot find symbol: variable & 41 & 57 & 57 & 57 & 57 \\ \cline{3-8}
  &                          & is already defined in class & 27 & 34 & 34 & 34 & 34 \\ \cline{3-8}
  &                          & cannot find symbol: method & 29 & 33 & 33 & 33 & 33 \\ \cline{3-8}
  &                          & incompatible types & 11 & 12 & 12 & 12 & 12 \\ \cline{3-8}
  &                          & illegal parenthesized expression & 5 & 9 & 9 & 9 & 9 \\ \cline{3-8}
  &                          & \begin{tabular}[c]{@{}l@{}}cannot be referenced from a\\ static context\end{tabular} & 5 & 4 & 4 & 4 & 4 \\ \cline{3-8}
  &                          & cannot assign a value to final variable & 4 & 6 & 6 & 6 & 6 \\ \cline{3-8}
  &                          & Illegal static declaration in inner class & 1 & 1 & 1 & 1 & 1 \\ \cline{3-8}
  &                          & \begin{tabular}[c]{@{}l@{}}method does not override or implement\\ a method from a supertype\end{tabular} & 0 & 1 & 1 & 1 & 1 \\ \cline{3-8}
  &                          & modifier static not allowed here & 0 & 1 & 1 & 1 & 1 \\ \cline{3-8}
  &                          & missing return statement & 0 & 0 & 0 & 0 & 0 \\ \cline{3-8}
  &                          & cannot be accessed from outside package & 0 & 0 & 0 & 0 & 0 \\ \cline{3-8}
  &                          & diamond operator is not supported in -source 6 & 0 & 0 & 0 & 0 & 0 \\ \cline{3-8}
  &                          & cannot be applied to given types & 0 & 0 & 0 & 0 & 0 \\ \cline{3-8}
  &                          & \textbf{Total} & \textbf{123} & \textbf{158} & \textbf{158} & \textbf{158} & \textbf{158} \\ \cline{2-8}
  & \multirow{2}{*}{JUnit}   & cannot find symbol: method & 4 & 3 & 3 & 3 & 3 \\ \cline{3-8}
  &                          & \textbf{Total} & \textbf{4} & \textbf{3} & \textbf{3} & \textbf{3} & \textbf{3} \\ \hline
\end{tabular}
}
\caption{Distribution of compilation errors by LLM, project, and run (with the specified long message split across two lines).}
\label{tab:compilation-errors-combined}
\end{table}

\rev{Moving forward with the attempts that did not report errors during our parsing, we also observe errors during the compilation process after applying and saving the changes as new commits (see Section \ref{sec:automatic-validation}).
Table \ref{tab:compilation-errors-combined} presents more fine-grained details about these errors.
First, we can observe that most errors are caused by hallucinations. 
Specifically, LLM-generated code snippets referenced symbols, like variables and methods, that are not available in the current context.
For example, consider the code snippet reported in Listing \ref{unavailable-symbol-gpt}, generated by \gptmini\ for the ANTLR4 project.
Such an example represents a \emph{semantic} error (name resolution), as the identifiers \texttt{data} and \texttt{offset} are not declared in the evaluated context, leading to the error ``cannot find symbol'' for both elements.
Similarly, we observe errors generated by DeepSeek, as reported in Listing \ref{missing-imports-deepseek} for the JUnit project.
Different from the previous example, this time, the error occurs due to missing static imports for JUnit assertions.
Methods like \texttt{assertNotNull} and \texttt{assertTrue} are static members of \texttt{org.junit.Assert}; without \texttt{import static org.junit.Assert.*;} (or using 
\texttt{Assert.assertNotNull(\dots)}), the compiler cannot resolve the symbols, and consequently, a failed attempt to compile the code is observed.
}

\begin{lstlisting}[style=javastyle,caption={Method referencing undeclared identifiers.}, label=unavailable-symbol-gpt]
public UUID execute() {
    long leastSigBits = toLong(data, offset);      // data, offset not in scope
    long mostSigBits  = toLong(data, offset + 4);  // data, offset not in scope
}
\end{lstlisting}

\begin{lstlisting}[style=javastyle,caption={JUnit test using missing assertions.}, label=missing-imports-deepseek]
public void testCountWithExplicitFilter() throws Throwable {
    {...}
    Result result = new JUnitCore().run(baseRequest.filterWith(include));
    assertNotNull("Result should not be null", result);
    assertTrue("Test should be successful", result.wasSuccessful());
}
\end{lstlisting}

\rev{Second, we have duplicated elements placed in the same scope, like two variables or two methods within the same method or class, respectively.
Third, we have cases related to incompatible types, like expecting a given type and receiving a different one.
The remaining cases were sporadic, not representing similar consistency across the different runs.
Overall, we can conclude that most of these errors are caused by the lack of context given to LLMs.
Considering the limited information provided, LLMs might start making assumptions, leading to the reported hallucinations.
Furthermore, we believe additional checks could be performed; for example, for \textit{duplicated methods}, our scripts could check whether a method with the same name exists in the target class. 
If so, our scripts could rename it and update associated calls, or just feed such information and prompt the LLM for an updated solution.
}

\subsection{Improving Code Assistant Tools}

\rev{As previously stated, we believe that our findings could further improve or assist current code assistant tools, like WindSurf and Cursor.\footnote{https://windsurf.com/, https://cursor.com/en}
First, based on our analysis of errors faced by the LLM-generated code snippets, we believe these assistant tools could be aware of possible errors and consequently, provide additional support to deal with them. 
For example, for \textit{cannot find symbol} errors, extracting further information about the error, like the name of the variable or method, and then further prompting the LLM to produce a new solution. 
For \textit{incompatible types} errors, we believe different motivations could take place. For example, hallucinations, as previously discussed, or due to different external dependencies.
For the former reason, a similar approach adopted for unavailable symbols could be enough.
For the latter, additional details could be provided, for example, informing about the version of the external dependencies adopted for the project.
}

\rev{Considering the diversity of prompts and refactoring types investigated here, we believe some insights can be gained.
First, for regular types of refactorings conventionally applied by developers supported by IDEs, like \textit{Extract method} or \textit{variable}, LLMs are good at doing them. 
However, for more unconventional cases, as investigated by our sample of Benchmark's scenarios, our results for real scenarios exploring \textit{rule-based} prompt reported good results.
We believe that such a constrained, precise, and direct way to perform the steps, while informing the expected outputs, could be used for these cases.
}

\rev{In the same way, regarding the implications for researchers, a possible future work could explore the development of autonomous systems capable of performing continuous refactoring in real-time as developers work on their codebases. 
These agents, which could be integrated directly into IDEs, could autonomously detect refactoring candidates without requiring explicit prompts. 
Additionally, they could also explore the benefits of integrating LLMs with human oversight in the refactoring process. 
Such a \textit{human-in-the-loop} approach would help ensure that the refactored code maintains its intended meaning and functionality, while addressing any issues related to semantic accuracy. 
We believe that involving humans could improve the transparency of the model’s decision-making, making it easier to understand how the LLM arrives at its conclusions and ensuring the quality of the refactorings in real-world applications.}

\subsection{Exploring LLMs-as-a-Judge}
To evaluate the potential of large language models in judging the correctness of applied refactorings, we adopt the LLM-as-a-Judge paradigm. This choice is motivated by recent studies that successfully employed LLMs for similar evaluation tasks \citep{zheng2023judging, gu2024survey}.
\rev{To mitigate potential bias between LLM generation and evaluation, we employ different models for these two roles. 
For instance, outputs produced by \textit{\gptmini} are evaluated by \textit{DeepSeek}, and vice versa. For each refactoring type, we prompt the evaluator LLM to assess whether the corresponding implementation is correct or not. 
The prompt includes the refactoring type along with the before and after code snippets, as illustrated as follows.
The reported prompt is based on the \texttt{zero-shot} prompt (see Section \ref{sec:llm_prompting}), as this approach yielded results comparable to those obtained with alternative prompting strategies.
}

\begin{tcolorbox}[myinstructionbox, title=Prompt for Checking the Implementation of LLM-generated Refactorings]
\textit{Given this initial code: \texttt{\$<initial\_code>\$}}

\vspace{0.5em}
\textit{the \texttt{\$<refactoring\_name>\$} was attempted and this was the final result:}
\vspace{0.5em}

\texttt{\$<final\_code>\$}
\vspace{0.5em}

\textit{Answer \texttt{1} if it was a success, \texttt{0} if it was a failure, and nothing else.}
\end{tcolorbox}

\rev{To assess the correctness of the LLM judgments, we rely on the human evaluation, as detailed in Section~\ref{sec:manual-validation}, and use it as the ground truth for our analysis.
For the refactorings performed by \gptmini\ (as judged by DeepSeek), the sample was highly imbalanced: DeepSeek labeled 94\% of the refactorings as successful, compared to 80\% by humans. This skew inflated chance agreement, yielding high raw agreement (77.7\%) but a very low Cohen’s Kappa (0.047). Positive agreement was strong (0.87), yet negative agreement was almost absent (0.14). PABAK, which adjusts for prevalence effects, indicated a more moderate agreement (0.555).
For the refactorings performed by DeepSeek (as judged by \gptmini), the distribution was more balanced: \gptmini\ labeled 70\% of refactorings as successful, while humans marked 50\%. With less skew, agreement improved: raw agreement of 70.7\%,  $\kappa = 0.345$ (fair), with positive (0.79) and negative (0.50) agreement both meaningful. Here, PABAK was 0.415, also consistent with moderate agreement.}

\rev{Taken together, these findings indicate that LLM judges tend to over-accept refactorings, and that imbalance strongly affects  Cohen's Kappa ($\kappa$). Reporting complementary measures such as positive/negative agreement and PABAK, alongside raw accuracy, provides a more comprehensive picture of reliability.
Finally, although our investigation focused exclusively on the \texttt{zero-shot} prompt, future work could explore alternative prompting strategies to assess whether they enable LLMs to provide more accurate judgments.}

\section{Threats to validity}\label{sec:threats}

This study is subject to several potential threats that could influence the results presented here. In the following section, we identify these threats and discuss the strategies we employed to mitigate their impact.

\textit{Construct to Validity.} To evaluate the correctness of the refactorings performed by LLMs on the Benchmark's scenarios, we conducted a manual analysis. To minimize potential bias, two researchers independently assessed the refactorings and then discussed any disagreements to reach a consensus. As previously mentioned, we observed high Cohen's Kappa coefficients, indicating strong agreement between the reviewers. Additionally, we explored the use of LLMs as judges, as suggested by Zhao et al. (\citeyear{zhao2024codejudge}).
\rev{As observed, LLM judges tended to classify a large majority of cases as successful; this skew strongly influenced chance-corrected measures such as Cohen’s kappa, resulting in artificially low values despite relatively high raw agreement. 
To mitigate this, we complemented kappa with positive/negative agreement and PABAK, which are more robust to prevalence effects. 
Another potential threat is that we relied on only two LLMs (\gptmini\ and DeepSeek) as both producers and judges of refactorings; outcomes may vary with other models or domains. 
Finally, while human annotations were used as ground truth, they remain subject to interpretation and possible error, which could also affect the reliability results.}

To assess the semantic impact of the generated refactorings on Real Scenarios, we relied on running the test suite associated with the target project. However, if the project's test suite is inadequate, it may fail to detect regressions, potentially leading to inaccurate conclusions regarding the correctness of the refactorings. To mitigate this risk, we selected popular projects known for having robust test suites. Additionally, we acknowledged the potential for flaky tests to introduce variability in the test results. To address this concern, we executed the tests five times for each scenario and observed no instances of test flakiness.

\rev{One of the threat to the validity of instructions arises from the examples used in Two-shot Learning strategy. The code examples provided in Fowler's book (step 6 in Figure \ref{fig:example-refactoring}) are written in JavaScript, whereas our datasets focus on Java projects. We intended to use these examples as part of the Two-Shot instruction; however, we chose not to translate the code snippets into Java. Instead, we instructed the LLMs within the context of the prompt to generate their outputs in Java.
This decision could potentially introduce bias, as LLMs might disregard the instruction, since the examples in the instructions are in JavaScript, and LLMs might therefore generate their output in the same language.
However, we did not observe cases like this, showing that the LLMs followed the language of input code to generate the final output.}
In designing our Rule-based instruction, we did not establish rules for all the refactoring types under analysis, since the rules were derived from an older version of Fowler’s catalog that was used in the Ref-Finder tool. As a result, 15 refactoring types were not included while applying Rule-based Learning that may impact the overall performance while applying this instruction. 

Another potential threat arises from the choice of quality metrics used to assess code after refactoring. Although the metrics selected in our study belong to a set of well-established measures that have been widely used in previous research on refactoring, they may not fully capture all aspects of code quality.

\textit{Internal Validity.} Our results may be subject to bias due to the non-deterministic behavior of LLMs. To mitigate this, and in line with previous studies, we prompted each LLM multiple times (five) for each refactoring scenario to reduce the impact of this variability. Another potential threat to internal validity is the presence of biases in our dataset arising from memorization, as discussed by Carlini et al. \citep{carlini2022quantifying}. Given that we selected Benchmark and real-world refactoring scenarios, there is a possibility that these scenarios were encountered during the training of the LLMs, which could influence the results. However, our results on the CodeBLEU metric reduce the likelihood of memorization issues, as the refactored code generated by LLMs shows low similarity with the ground truth, with CodeBLEU scores below 0.5.

\rev{In addition, since the code snippets collected from Fowler’s book are independent examples intended to illustrate refactoring concepts, one possible solution instead of manual evaluation would be to automatically generate test cases for each snippet. However, we were unable to generate test cases for all code snippets. For instance, in one of the cases of \textit{Change Function Declaration}, where a function was renamed from \texttt{circum} to \texttt{circumference}, only the method signature was available, without information about the required parameters or method body, making automatic test generation not feasible. For this reason, the correctness and semantic preservation of the code in the \textit{Benchmark Scenario} dataset were assessed through manual validation.}

\textit{External to Validity.} Our findings are limited to the context of a single programming language, Java. While Java is widely used, restricting the study to this language may result in biased conclusions, as different programming languages have unique syntax, rules, and assistive tools. Similarly, we evaluated the capabilities of two LLMs, both of which are commonly used for various tasks in software engineering. Conducting this study with other models or employing advanced adaptation techniques such as fine-tuning could yield different results.

\section{Conclusion} \label{sec:conclusion}
\rev{This study examined the ability of LLMs to perform code refactoring beyond refactoring types with simple transformation, considering both the breadth of refactoring types and the role of different instruction strategies inpired by Martin Fowler's guideline. Our findings show that LLMs can generate refactorings that preserve semantics and, in some cases, improve code quality metrics such as complexity depending on the provided instructions. Rule-based Learning, inspired by automated refactoring tools, guided \gptmini\ more effectively, while DeepSeek benefited from minimal guidance under Zero-Shot learning that includes only the name of refactoring type. We also found that Objective-based Learning, which only describe the overall goal of refactoring, can lead to improvements in code quality even when the intended transformation is not precisely applied. Overall, while DeepSeek outperformed \gptmini\ in the number of successful refactorings, both models exhibited limitations in semantic and quality preservation.}

\rev{While this study offers initial insights into the potential of LLMs for code refactoring, several limitations leave many avenues for future research. Future work should extend this analysis to additional models, programming languages, and richer contextual settings, while covering a wider aspects of code quality improvements.}

\section*{Acknowledgements}
\label {acknowledgements}
We thank the anonymous reviewers for their valuable comments on improving an earlier version of this paper. 

\section{Declarations}

\subsection{Funding}
This work is funded by the following organizations and companies: Fonds de Recherche du Quebec (FRQ), Natural Sciences and Engineering Research Council of Canada (NSERC), the Canadian Institute for Advanced Research (CIFAR), and the Canada Research Chairs Program. However, the findings and opinions expressed in this paper are those of the authors and do not necessarily represent or reflect those of the organizations/companies.

\subsection{Ethical approval} 
Not applicable.

\subsection{Informed consent} 
Not applicable.

\subsection{Author Contributions}
\textbf{Yonnel Chen Kuang Piao:} Conceptualization, Data Curation, Formal Analysis, Investigation, Methodology, Visualization, Writing– Original Draft, and Writing– Review \& Editing.
\textbf{Jean Carlors Paul:} Conceptualization, Data Curation, Investigation, Methodology, Writing– Original Draft, and Writing– Review \& Editing.
\textbf{Leuson Da Silva:} Conceptualization, Methodology, Writing– Original Draft, and Writing– Review \& Editing.
\textbf{Arghavan Moradi Dakhel:} Conceptualization, Methodology, Writing– Original Draft, and Writing– Review \& Editing.
\textbf{Mohammad Hamdaqa:} Supervision, Validation, and Writing– Review \& Editing.
\textbf{Foutse Khomh:} Project Administration, Resources, Supervision, Validation, and Writing– Review \& Editing.

\subsection{Data Availability}
\label{sec:data_availability}
To promote open science and facilitate reproducibility, we make all our artifacts available to the community.
This includes the scripts used for evaluation and the LLM-generated code snippets during our study, available in our Online Appendix (\citeyear{appendix}).

\subsection{Conflicts of Interest}The authors declare that they have no known competing financial interests or personal relationships that could have appeared to influence the work reported in this paper.

\subsection{Clinical trial number} 
Not applicable.

\bibliographystyle{apalike}
\bibliography{main}

\end{document}